\documentclass[a4paper,UKenglish,cleveref, autoref, thm-restate]{oasics-v2021}
\usepackage{xspace}
\usepackage{booktabs}
\nolinenumbers

\definecolor{darkkhaki}{rgb}{0.74, 0.72, 0.42}

\newcommand{\name}{SwiftQueue\xspace}

\newcommand{\eg}{{\it e.g.,}\xspace}
\newcommand{\ie}{{\it i.e.,}\xspace}

\definecolor{backcolour}{rgb}{0.96,0.96,0.96}
\definecolor{codegray}{rgb}{0.5,0.5,0.5}
\definecolor{deepblue}{rgb}{0,0,0.6}
\definecolor{deepred}{rgb}{0.6,0,0}
\definecolor{deepgreen}{rgb}{0,0.5,0}
\lstdefinestyle{mystyle}{
    backgroundcolor=\color{backcolour},   
    commentstyle=\color{codegreen},
    morekeywords={self, True},
    keywordstyle=\color{deepblue},
    numberstyle=\tiny\color{codegray},
    emph={MyClass,__init__,EncodingType,Image},
    emphstyle=\color{deepred},
    stringstyle=\color{deepgreen},
    basicstyle=\ttfamily\footnotesize,
    breakatwhitespace=false,         
    breaklines=true,                 
    captionpos=b,                    
    keepspaces=true,                 
    numbers=left,                    
    numbersep=5pt,                  
    showspaces=false,                
    showstringspaces=false,
    showtabs=false,                  
    tabsize=1
}

\title{\name: Optimizing Low-Latency Applications with Swift Packet Queuing} %TODO Please add

\titlerunning{\name: Optimizing Low-Latency Applications with Swift Packet Queuing} %TODO optional, please use if title is longer than one line

\author{Siddhant Ray\footnote{Corresponding author}}{University of Chicago, USA \and \url{https://siddhant-ray.github.io/} }{siddhantray@uchicago.edu}{https://orcid.org/0000-0003-0265-2144}{}

\author{Xi Jiang}{University of Chicago, USA \and \url{https://chasexj.github.io/} }{xijiang9@uchicago.edu}{https://orcid.org/0000-0003-1652-8419}{}

\author{Jack Luo}{University of Chicago, USA }{luoqiansu@uchicago.edu}{https://orcid.org/0009-0003-1255-6364}{}

\author{Nick Feamster}{University of Chicago, USA \and \url{https://people.cs.uchicago.edu/~feamster/} }{feamster@uchicago.edu}{https://orcid.org/0000-0001-9315-5201}{}

\author{Junchen Jiang}{University of Chicago, USA \and \url{https://people.cs.uchicago.edu/~junchenj/} }{junchenj@uchicago.edu}{https://orcid.org/0000-0002-6877-1683}{}

\authorrunning{S Ray et al.} %TODO mandatory. First: Use abbreviated first/middle names. Second (only in severe cases): Use first author plus 'et al.'

\Copyright{Siddhant Ray and Xi Jiang and Jack Luo and Nick Feamster and Junchen Jiang} %TODO mandatory, please use full first names. LIPIcs license is "CC-BY";  http://creativecommons.org/licenses/by/3.0/

 %TODO mandatory; please add comma-separated list of keywords

\category{} %optional, e.g. invited paper

\relatedversion{} %optional, e.g. full version hosted on arXiv, HAL, or other respository/website
\EventEditors{Katerina J. Argyraki and Aurojit Panda}
\EventNoEds{2}
\EventLongTitle{1st New Ideas in Networked Systems (NINeS 2026)}
\EventShortTitle{NINeS 2026}
\EventAcronym{NINeS}
\EventYear{2026}
\EventDate{February 10, 2026}
\EventLocation{Virtual Conference}
\EventLogo{}
\SeriesVolume{139}
\ArticleNo{24}
\begin{document}

\maketitle

% \begin{abstract}
% Lorem ipsum dolor sit amet, consectetur adipiscing elit. Praesent convallis orci arcu, eu mollis dolor. Aliquam eleifend suscipit lacinia. Maecenas quam mi, porta ut lacinia sed, convallis ac dui. Lorem ipsum dolor sit amet, consectetur adipiscing elit. Suspendisse potenti. 
% \end{abstract}
\begin{abstract}

\emph{Low Latency, Low Loss, and Scalable Throughput} (L4S), as an emerging router-queue management technique, has seen steady deployment in the industry. 
An L4S-enabled router assigns each packet to the queue based on the packet header marking. Currently, L4S employs {\em per-flow} queue selection, \ie all packets of a flow are marked the same way and thus use the same queues, even though each packet is marked separately. 
However, this may hurt tail latency and latency-sensitive applications because transient congestion and queue buildups may only affect a fraction of packets in a flow. 
 
We present {\em \name}, a new L4S queue-selection strategy in which a sender uses a novel {\em per-packet} latency predictor to pinpoint which packets likely have latency spikes or drops. 
The insight is that many packet-level latency variations result from complex interactions among recent packets at shared router queues. 
Yet, these intricate packet-level latency patterns are hard to learn efficiently by traditional models. 

Instead, \name uses a custom {\em Transformer}, which is well-studied for its expressiveness on sequential patterns, to predict the next packet's latency based on the latencies of recently received ACKs. Based on the predicted latency of each outgoing packet, \name's sender dynamically marks the L4S packet header to assign packets to potentially different queues, even within the same flow. 
Using real network traces, we show that \name is 45-65\% more accurate in predicting latency and its variations than state-of-art methods. Based on its latency prediction, \name reduces the tail latency for L4S-enabled flows by 36-45\%, compared with the existing L4S queue-selection method.

\end{abstract}
\section{Introduction}
\label{sec:intro}

\begin{figure}[ht]
    \centering
    \includegraphics[width=0.6\textwidth]{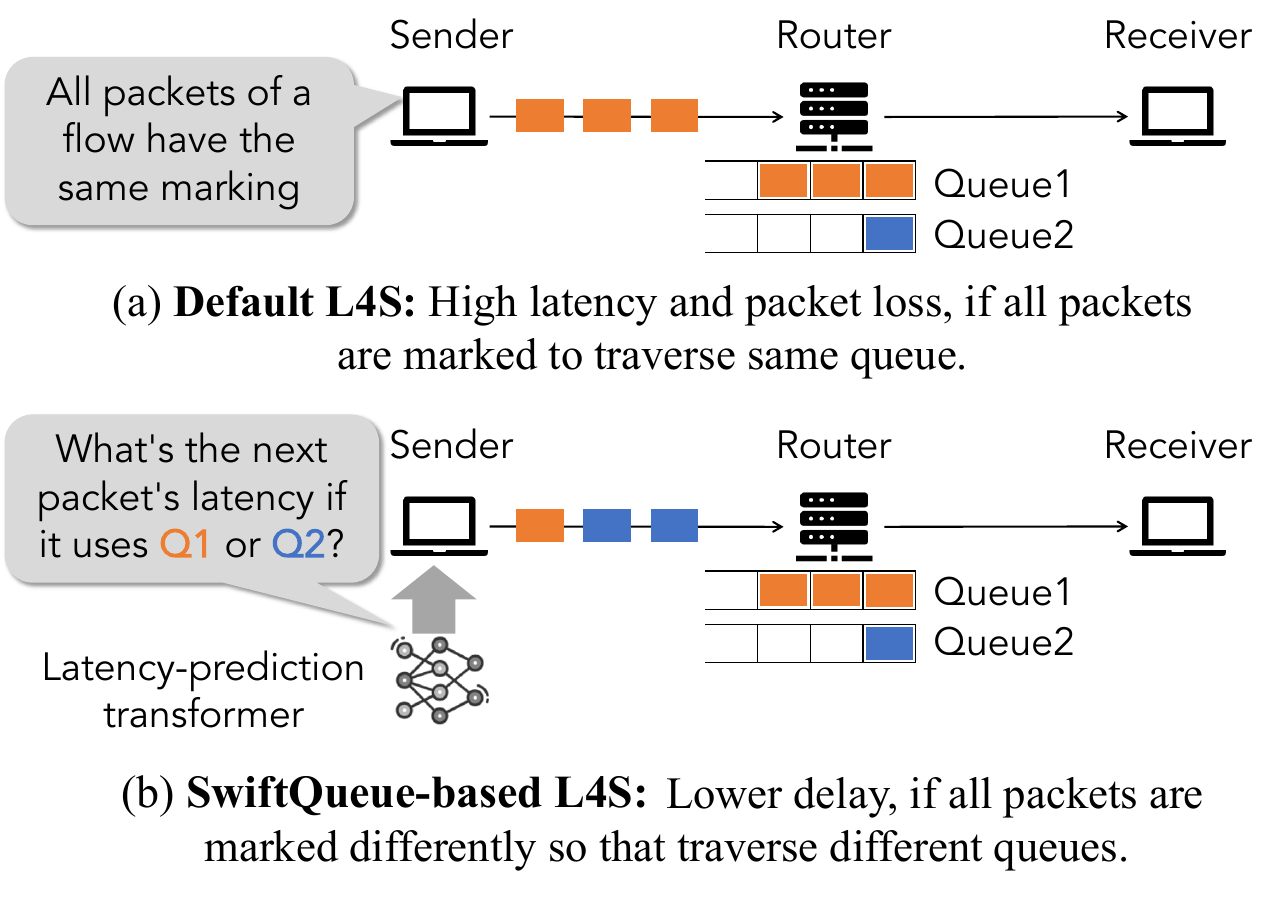}
    \caption{To reduce tail latency, \name predicts latency per packet and choose the appropriate queue for each packet.}
    \label{fig:motivation}
\end{figure}

Internet Service Provider (ISP)-supported QoS enhancement in wide area networks has been long-awaited. 
Fortunately, the gradual yet steady deployment of L4S in ISPs over the last few years enables sender-driven queue selection on the network routers~\cite{comcast_l4s, l4s_deployment_1, l4s_deployment_2, l4s_deployment_3, l4s_deployment_4}.
The sender marks the outgoing packet headers of L4S-enabled flows, while the L4S-enabled routers along the path will use different queues for these packets depending on the marked headers.
% This allows the sender to decide which queues a packet or a group of packets would traverse over a path. 
With this new capacity, the sender could mark the packets in a way such that packets that would otherwise experience high latency can now go through alternative lower-latency queues, potentially reducing the tail end-to-end latency for low-latency applications.
% . This, in turn, can reduce the tail packet latency, which is critical for latency-sensitive applications, and also increase bandwidth utilization in the network.

As with many previous sender-driven queue selection techniques
(\eg~\cite{diffserv,efphb,afphb}), L4S marks {\em all packets} of a flow in
the same way, \ie all packets in a flow are assigned to the same queue.
Yet, even within a flow, it is beneficial to select queues differently for
different packets, as the latency of a router queue can temporarily increase
or decrease significantly. % due to localised congestion in a short time window. 
Drastic changes in latency may occur due to inter-packet interactions on a \emph{shared network} in a short time window~\cite{10.1145/149439.133097}, which can cause sudden changes in the network state. 
Realizing the full potential of L4S's sender-driven queue selection thus requires predicting precisely when latency spikes and drops happen at the per-packet level.

Predicting sudden, per-packet latency changes is a longstanding challenging problem~\cite{harsha_latency, matthew_latency, kevin_latency,global_latency}; and most congestion control and queuing schemes only \emph{react} to packet-level latency changes.
% instead of proactively preventing them.  
Inspired by the Transformer’s success in other time-series domains (\eg natural languages)~\cite{vaswani2017attention}, we 
% revisit the problem of packet-level latency-change prediction. 
view the sequence of packets and their latencies as a token sequence, and ask a simple question: 
{\em can Transformers learn to predict packet latency accurately enough to allow L4S to proactively reduce tail latencies for low-latency applications?}
% \emph{can Transformer learn to predict, before sending the next packet out, whether the packet will experience significantly different latency than the preceding packets?}

This paper shows that Transformers when trained appropriately, can predict sharp latency changes much more accurately than prior methods, even when tested on network traces unseen in training. 
This stems from the observation that while some latency variations are indeed due to exogenous factors (\eg user clicks), not all changes are. It is possible to have some learnable events in the network indicated by signals embedded in the latencies of the past packets. 
For instance, video applications (\eg Netflix or Zoom) can have periodic packet bursts of video chunks or frames within a session, and these flows can have some learnable patterns for their latencies~\cite{10.5555/334590.2813153,10.1145/570681.570689}. 
For another example, the interaction of packets from running flows using certain congestion control logic can result in learnable latency patterns (\eg queue buildup and back-off in TCP Cubic~\cite{tcp_cubic}).
Of course, the real latency patterns can be more complicated than these simple examples.
% We use these examples here for ease of understanding, but the real latency patterns can be more complicated.

Learning to extract signals from past packets' latencies has been challenging 
% due to the limited abilities of 
for traditional time-series models. % to capture dependencies across packet latencies. 
Specifically, simple models (\eg EWMA~\cite{ewma}, ARIMA~\cite{arima}) can learn seasonality and smooth past values, but they fall short of capturing more complex and finer-grained inter-packet dependencies~\cite{zeng2023financialtimeseriesforecasting}.
% as they rely on seasonality and smoothing over the past values. 
Deep learning models offer better learning and expressiveness, but training and updating these models can be slow and data-intensive~\cite{khanal2024domainadaptationtimeseries,chang2024llm4tsaligningpretrainedllms}. 
% but to cope with the variation of competing flows and applications, the prediction model needs to be updated frequently~\cite{khanal2024domainadaptationtimeseries,chang2024llm4tsaligningpretrainedllms}. 
The Transformer's unique advantage comes from its parallel \emph{attention} mechanism, which simultaneously captures complex sequential patterns within a long context window and allows faster training and convergence than pre-Transformer neural networks, \eg Recurrent Neural Networks (RNNs), Long-Short Term Memory Networks (LSTMs)~\cite{lstm}
% . These are unique advantages, compared to both classical ML techniques and pre-Transformer neural networks, \eg Recurrent Neural Networks (RNNs), Long-Short Term Memory Networks (LSTMs)~\cite{lstm} for time-series predictions. 
For these reasons, Transformers have shown great promise in general time-series predictions~\cite{nie2023time, zhou2021informer, zhou2022fedformer, wu2022autoformer}.

Inspired by this, we present {\em \name}, a new L4S queue-selection system driven by a {\em custom Transformer} for per-packet latency prediction.
% a Transformer architecture, as a reliable latency prediction-driven queue selector at the packet level. 
\name has two components: 
(1) a Transformer that predicts next outgoing packet's latency based on information (latencies and queue selections) of recent packets from the same sender, and 
(2) a queue-selection logic that uses the Transformer-based latency prediction to select the queue for outgoing packets in an L4S flow and marks the packet headers accordingly so that L4S-enabled routers will put the packet in the corresponding queues. 
As illustrated in Figure~\ref{fig:motivation}(b), since \name assigns queues per packet rather than per flow, 
it can reduce tail latency. %, especially for low-latency applications. 
% Figure~\ref{fig:motivation} shows how \name's controller would reassign packets into individual queues.

% While the use of Transformers in networking is not new~\cite{10.1145/3563766.3564104, guthula2023netfound,m3,telemetry}, 
\name presents the first effort to adapt Transformers for {\em packet-level} latency prediction. 
Our custom Transformer focuses on neighborhoods of sharp changes and learns to predict the latency of future packets based on the timeseries of past latencies and packet information.
% network's effects embedded in past values. 
Unlike prior uses of Transformers in networking~\cite{10.1145/3563766.3564104, guthula2023netfound,m3,telemetry}, which focuses on {\em flow-level} estimates or classification, our goal requires learning latency patterns due to {\em cross-flow} interactions at the packet level and making fast prediction per packet. 
% In previous work, inference is done at the flow-level (flow-level estimates) or the model focuses on making decisions for the entire flow (flow-classification).

\paragraph*{Contributions} In short, we make the two contributions:

\begin{enumerate}
    \item First, we show that our Transformer predictor, which is continuously fine-tuned using past packet latency values, can accurately predict latency spikes or drops 45-65\% more accurately than the state-of-the-art methods. Moreover it can make per-packet latency prediction fast enough for a 200-Mbps link by making batched predictions. % of the next set of latency values.
    \item Second, we show that, using \name, a sender can significantly reduce the tail latency by 36-45\%, by first predicting the latency of each outgoing packet in the L4S flow if it goes through the L4S queues or classic queues and then marking each packet for queue selection such that its predicted latency is minimized.
\end{enumerate}

% \section{Motivation and Related Work}
\section{Background}
\label{sec:motivation}

\subsection{Queue selection and management}
\label{subsec:usecases}

\begin{figure}[t]
    \centering
    \includegraphics[width=0.75\textwidth]{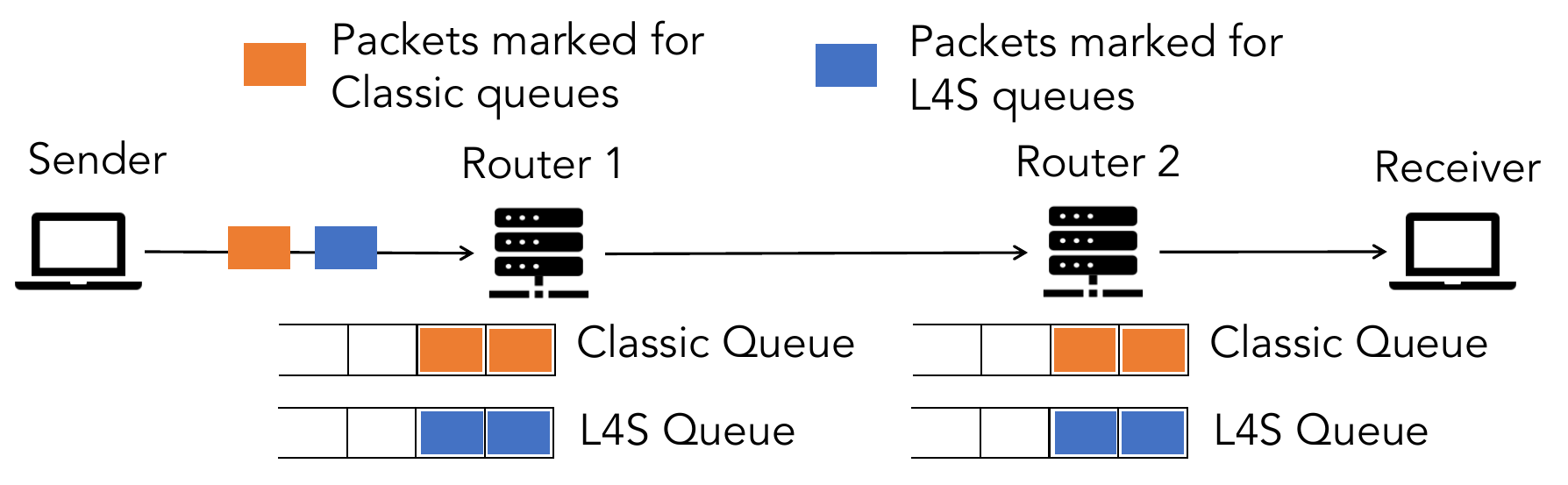}
    \caption{L4S by default marks all packets from the flow into the same queues, and once marked, packets traverse the chosen queues at all hops.}
    \label{fig:l4s_bg}
    \end{figure}

\paragraph*{L4S dual queue}
    L4S is already being rolled out for deployment at scale~\cite{l4s_deployment_1, l4s_deployment_2, l4s_deployment_3, l4s_deployment_4} and is expected to be the dominant solution for reducing queuing latency for low-latency applications (\eg cloud gaming) and networks (low-latency 5G/ISP).
    The L4S Dual Queue~\cite{l4s, l4s2, l4s3} is developed to mitigate bufferbloats and excessive queuing latency in networks. 
    With the L4S Dual Queue, the sender tags a packet header's ECN bit~\cite{ecn} to indicate if a packet should go through the ``Classic'' queues or the ``L4S'' queues in the L4S-enabled routers (Figure~\ref{fig:l4s_bg}). 
    % is placed in one queue, and the L4S traffic is placed in another. 
    To sort the traffic into each of the two queues correctly, the L4S dual-queue relies on packets being tagged appropriately. 
    % Classic traffic is tagged as Non-ECN whereas L4S traffic is tagged as ECN-capable, which ensures the non-interference with each other.
    This enables different rules for ECN marking and/or dropping in each of the queues.
    That said, currently, L4S senders only select queues per flow/application based on the application's nature (video or web) or user preference.

    % L4S uses a bit in the IP header as an explicit congestion signal, \ie the Explicit Congestion Notification (ECN) bit~\cite{ecn}. 
    % % Using ECN makes it possible to signal congestion early. 
    % Since an ECN congestion signal can be sent without paying the costs of dropping packets, this is one of the major strengths of L4S. 
    % % ECN signals allow the sender to react less dramatically to congestion signals, so the transmission rate can be fine-tuned quicker and more accurately. 
    % L4S relies on a DCTCP style algorithm named TCP Prague, which enables ECN support for signaling congestion.

    % % Purely enabling ECN in the network to support TCP Prague would doesn't work on the internet because other TCP versions (already deployed) could get a similar high and frequent amount of ECN markings or packet losses, and react by reducing their throughput too aggressively. It is necessary to apply a lower amount of drops or classic ECN marks to classic traffic, than you would for L4S traffic. Hence we need to separate L4S and Classic packets in different queues in order to achieve the desired low latency.

    % With the L4S Dual Queue, classic traffic is placed in one queue, and the L4S traffic is placed in another. This enables different rules for ECN marking and/or dropping in each of the queues. To sort the traffic into each of the two queues correctly, the L4S dual-queue relies on packets being tagged appropriately. Classic traffic is tagged as Non-ECN whereas L4S traffic is tagged as ECN-capable, which ensures the non-interference with each other.

\paragraph*{Active queue management (AQM)}
    Before L4S, AQM has already been an active area of study for decades to ensure the efficient co-existence of TCP flows. % in terms of latency reduction, fairness etc. 
    Several AQM methods have been proposed for optimising different aspects, including queuing latency reduction, such as Random Early Detection~\cite{red}, CoDel~\cite{codel} and FQ-CoDel~\cite{fqcodel}. Traditional AQM algorithms have been built to run in conjunction with TCP congestion control algorithms, which rely only on packet loss as a signal (as opposed to L4S, which uses ECN markings). 
    To handle the bursty nature of TCP, these AQM techniques are equipped with large data buffers to prevent excessive packet drops due to these bursts. However, bufferbloat arises when queue length grows~\cite{bufferbloat}, and
    % if the buffers are increasingly large, this also means the packets are queued in much deeper queues, leading to excessive queuing latency for packets, a case which causes \emph{bufferbloat}~\cite{bufferbloat}. 
    latency can often build up as a result of individual bufferbloats at multiple routers on the network path. As traditional AQM techniques do not allow per-packet queue selection, they are unsuitable for low-latency applications where latency spikes happen at the packet level.

\subsection{Related work}
\label{subsec:existing_work}

\paragraph*{Latency prediction}
Latency prediction in networks has been studied for decades. 
Most recent work focuses on estimating latency values at the flow level (\eg flow completion times or average packet latency in a flow) using heuristics~\cite{scalable_latency, simon} or deep learning~\cite{m3}. There has also been work on measuring latency for network performance estimation at scale~\cite{chronos, network_delay_changes, pingmesh}. Other works have focused on predicting latency from noisy and incomplete network measurements~\cite{latency_pred_adaptive}. Finally, latency prediction has been actively explored at the application level for video streaming~\cite{livenet,insitu}, HTTP web traffic~\cite{modern_web_traffic, predictive_prefetching}, cloud datastores~\cite{c3}, and CDN serving~\cite{anyopt}.

\paragraph*{Machine learning for network traffic analysis}
Recent advancements in machine learning and deep learning have significantly impacted network traffic analysis for various tasks, from conventional
anomaly detection and traffic classification~\cite{jiang2023ac,lotfollahi2020deep} to more complex behavior characterization~\cite{brown2023augmenting} and traffic content forecasting~\cite{jiang2023generative,jiang2024netdiffusion}.
Various approaches have been proposed, including conventional models like simple regression and tree-structured classifiers~\cite{jin2020swiftids,lai2020using,akem2023flowrest}, as well as advanced deep learning techniques such as recurrent neural networks (RNNs)~\cite{kim2018web,radford2018network}, convolutional neural networks (CNNs)~\cite{liu2023real} and diffusion models~\cite{sivaroopan2023netdiffus,jiang2024netdiffusion}.
These methods have shown promise in understanding patterns in network traffic data but often fall short in handling long-range dependencies and sharp changes in traffic patterns~\cite{aouedi2021performance,ismailaj2021deep}. 

Transformers have been proposed as a new class of models to reason about more complex sequential dependencies in a variety of fields. In the networking domain, the Transformer architecture has already shown promise in predicting flow-level behaviors~\cite{10.1145/3563766.3564104, guthula2023netfound,m3}, telemetry signals~\cite{telemetry}, and other time-series~\cite{wu2024netllm}. They have also been used recently to generate network traffic~\cite{netgpt, zhou2024netflowgenleveraginggenerativepretraining} and analyze traffic behaviour at the flow level~\cite{trafficgpt}. 
These architectures use decoder-based Transformers to generate packet headers with semantically correct header structure but often lack temporal inter-packet patterns. Predicting actual packet latency values is different from generating packet headers. We need to learn packet dependencies within flows and across flows from an endpoint, which the models above do not address yet.

% We summarize the following key findings:
In short, we summarize the background as follows:

\begin{enumerate}
    \item Prior sender-driven queue selectors, including L4S's header marking strategy, 
    % While L4S allows per-packet queue selection, currently, it 
    only selects queues per flow/application as it does not predict packet-level events.
    \item Transformers have shown promise in some networking problems but not for per-packet latency prediction.
\end{enumerate}

% We also only focus on queue selection and we don't adapt the sending logic of the congestion control algorithm running at the sender.

\section{Motivation}
\label{sec:challenges}

Inspired by the Transformer's promise, this work applies the Transformer to packet-level queue-selection in L4S.
Importantly, our work is focused on queue selection at the \emph{sender} and {\em without} changing the L4S routers or the sender's congestion control logic.

% \subsection{Packet-level queue selection is important}
\subsection{\hspace{-0.2cm}L4S queue selection: per-{\em packet} vs. per-{\em flow}}
% should vary with packets, not flows}
\label{subsec:queue_selection}

Mechanically, L4S AQM architecture allows the sender to mark each packet in a flow differently into different queues, but by default, all packets of a flow are either sent through the classic queues or the L4S queues.
This can be suboptimal. 
For example, if the network has more low-latency flows than classic traffic flows, L4S will, by default, put all packets from low-latency flows on the L4S queue on the bottleneck router, causing the L4S queue to build up and create additional queuing latency for low-latency flows.

% Prior AQM techniques don't allow packet-level queue selection by the senders. L4S AQM architecture supports marking packets into different queues at the sender but it doesn't have an mechanism to make this decision. Our goal is to enable sender-driven queue assignment using per-packet latency predictions, as there is a huge potential to reduce tail latency for flows, by focusing on the particular packets which have high latency due to queue buildup and congestion.

% For example, a possible scenario is when the network has several L4S-enabled flows, all of which require low-latency. The current number of concurrent non-L4S traffic flows could be much lower than the number of L4S flows. Under the standard L4S setup all packets from the L4S flows will be put on the L4S queue on the bottleneck router while all other packets would be put on the classic queue. This causes the L4S queue to buildup and create additional queuing latency.

Current L4S vendors pre-partition queues for classic and L4S traffic for a given link subscription. The L4S queue is designed to be much shorter as L4S latency-sensitive traffic has been measured around 10-20\% of the overall network traffic.\footnote{Data collected after conversation with a leading US telecom vendor deploying L4S on their network.} However due to these fixed size queue allocations, if latency-sensitive traffic becomes capacity seeking on the link even for a short window, packets which would experience high-latency can be switched to the classic queue. This prevents L4S flows from experiencing latency spikes which would adversely have affected application-level Quality-of-Experience (QoE).

This way, one could potentially reduce tail latency by focusing the use of L4S queues on only the packets with high latency due to transient queue buildup and congestion.
Such per-packet queue selection could particularly benefit the flows that have stringent tail-latency requirements and only share a part of the entire network path with the other flows.
% has significant benefits in network scenarios where we have strict latency requirements for a subset of L4S flows, which might only share a part of the entire network path with the other flows. 
For example, if we have some L4S flows at a given router, which has packets that would experience further downstream congestion (due to cross traffic), we can select a non-congested queue for those packets at the current router, minimizing the tail latency as a result of those packets. We evaluate this in detail in \S\ref{eval:ss3}.

With a per-packet latency predictor, we can predict which packets in the low-latency flows would experience high latency and then mark them to let the router put just these packets into a different, lower-latency queue (classic queue) than other packets in the same flow, which reduces the tail latency. 
Meanwhile the current L4S queue would have been serviced and we can continue to put the next incoming packets back on the L4S queue after its queue length becomes low. 
% , to maintain the low-latency requirements for all packets. 

\subsection{\hspace{-0.2cm}Latency prediction needs expressive models}
\label{subsec:perpacket_model}

To enable packet-level queue selection, per-packet latency prediction is essential, especially when latency spikes and drops happen. 
% For queue selection to minimize latency, we need to predict the moments when latency spikes and drops happen in the time-series of packets. 
Classical ML models have been shown to perform poorly at capturing sudden variations in time-series~\cite{deepl_classic}. Neural network models are superior at learning complex feature relationships in time-series data as they have much greater memorization and reasoning. 
% In general time-series predictions, the Transformer architecture has shown promise as a transformative tool~\cite{??}.

\begin{figure}[t]
    \centering
    \includegraphics[width=0.68\textwidth]{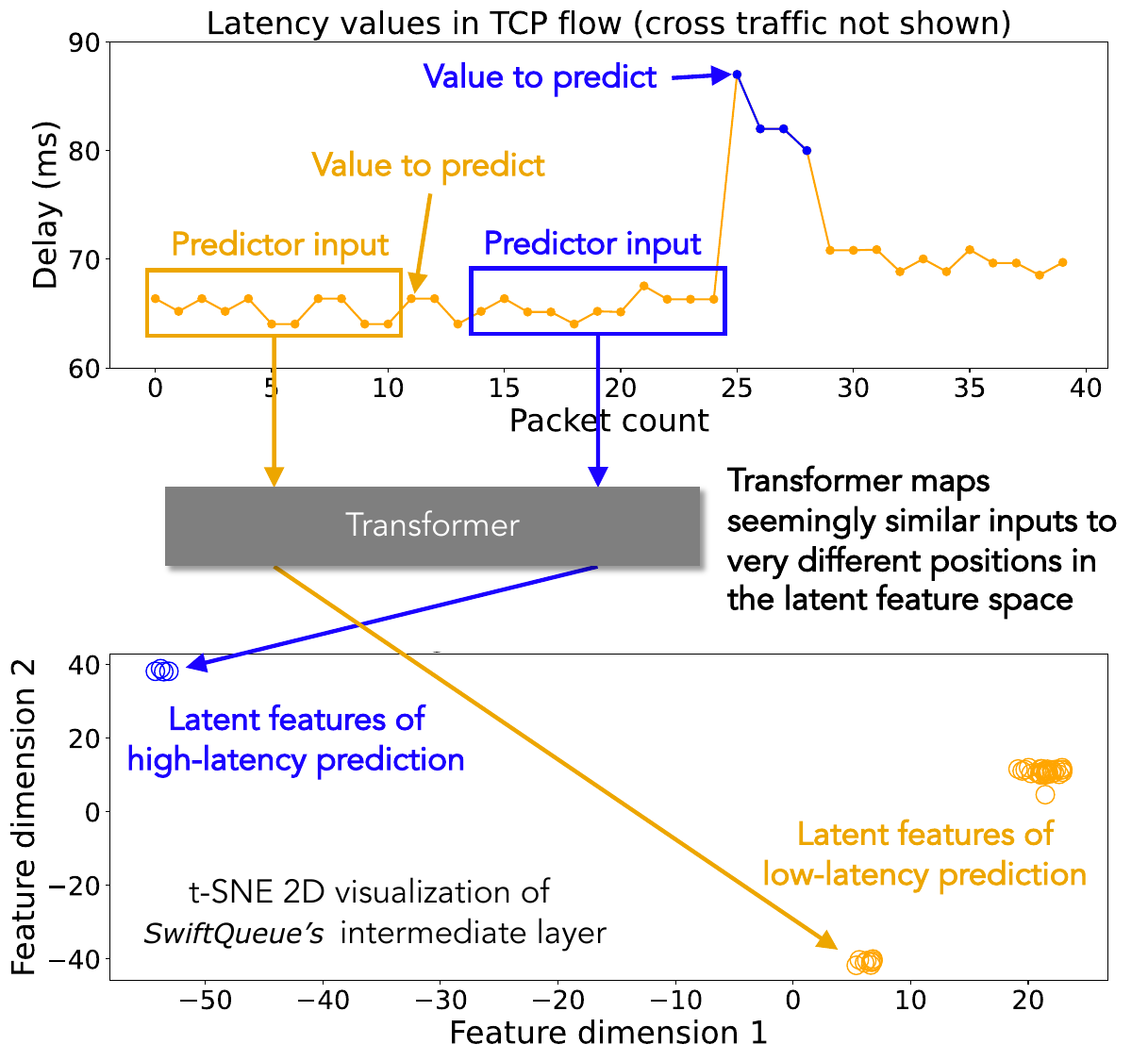}
    \caption{A Transformer maps the inputs of low-latency packets and those of high-latency packets to separate clusters.}
    \label{fig:latency_analysis}
\end{figure}

% \mypara{Transformers for time-series predictions}
% Transformers~\cite{vaswani2017attention} initially introduced in the field of natural language processing (NLP) have revolutionized various domains due to their capability to handle sequential data efficiently. 
The core innovation of Transformers is the \emph{self-attention} mechanism~\cite{vaswani2017attention}, which allows the model to weigh the importance of different elements (tokens) in a sequence when making predictions. 
The self-attention mechanism works by computing a weighted sum of input elements, where the weights are determined by the relevance of each element to the current prediction.
This allows focusing on different parts of the input sequence in an adaptive manner, enhancing its ability to capture dependencies and interactions.

Transformers have shown promise as a superior tool for general time-series prediction tasks~\cite{li2019enhancing,lim2021temporal,nie2023time, zhou2021informer, zhou2022fedformer, wu2022autoformer}, leveraging the ability of the attention mechanism to capture complex temporal patterns much better than other deep learning models.
% These models incorporate modifications such as time-based positional encodings to handle the sequential nature of time-series data. 
Transformer's \emph{attention} mechanism offers greater memorization of past interactions~\cite{memorizingtransformers}. While other deep learning models like LSTMs and RNNs have been used for time-series prediction, they 
% don't perform as well as Transformers. They 
suffer from well-studied issues of \emph{memory limitations}~\cite{Illium_2022,staudemeyer2019understandinglstmtutorial} and \emph{vanishing gradients}~\cite{li2018independentlyrecurrentneuralnetwork, pascanu2013difficultytrainingrecurrentneural} which reduces their prediction accuracy and training speed.

To showcase that a Transformer, even at a small size, could capture the complex patterns before latency changes, Figure~\ref{fig:latency_analysis} visualizes the intermediate feature space of different Transformer inputs. 
This Transformer, which has 2 layers (will be described in more detail soon), predicts a given packet's latency based on recent packets' latencies as input.  The intermediate feature, a $64\times64$ tensor, is visualized by first mapping the tensor to a two-dimensional space using t-SNE~\cite{tsne}.

As an example, we consider two particular different packet latencies under prediction. 
% packets under prediction have very different actual latencies. 
While their inputs may seem similar (the blue box and the yellow box), their features produced by the Transformer fall in well-separated clusters. 
Specifically, for the {\em very first} packet after a latency spike (the blue point), the Transformer already maps them to the correct feature region, even if significantly increased or decreased delay is not visible in their inputs. 
In other words, even though their inputs have little noticeable difference to packets prior to the spike/drop, the Transformer can tell the difference from the distribution embedded in the values!

% In Figure \ref{fig:latency_analysis} we show a sequence of latency values from a single TCP Cubic flow (from a set of flows) simulated in NS-3. We see an intermediate layer of \name's predictor (introduced in Section \ref{sec:design}) featurizes high latency packets and low latency packets very differently in its high-dimensional representation. We use a popular visualization technique, t-SNE~\cite{tsne} to visualize these high dimensional feature representations by reducing into 2-dimensional features. We see well-separated clusters for different latency value neighbourhoods which shows the Transformer can learn separation of different latency values in packet sequences.

\section{\name Design}
\label{sec:design}

\begin{figure*}[t]
    \centering
    \includegraphics[width=\linewidth]{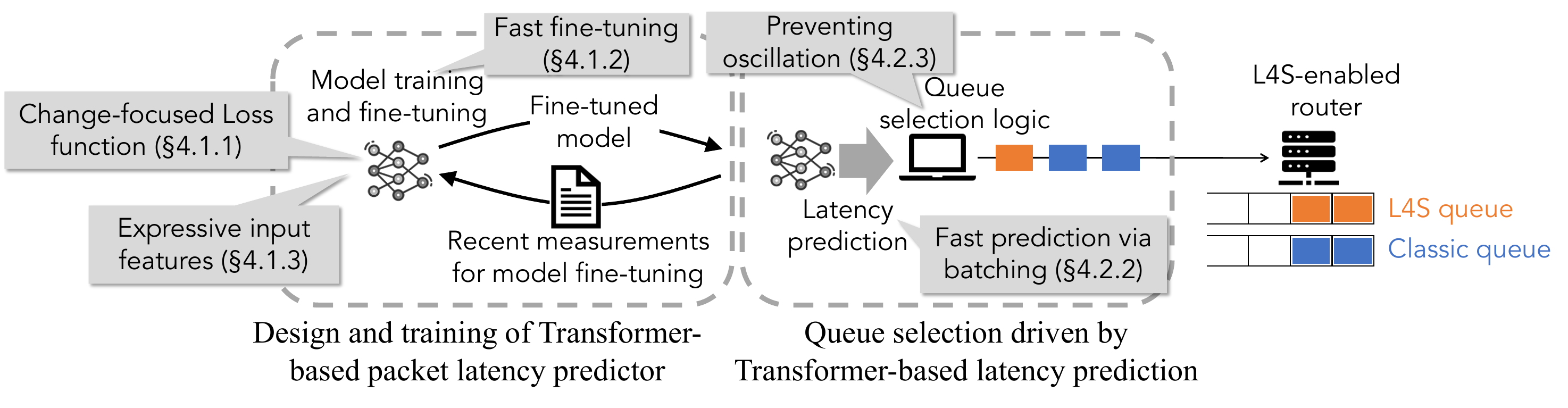}
    \caption{\name consists of a custom Transformer latency predictor and a prediction-driven per-packet queue selector.}
    \label{fig:overview}
\end{figure*}

Now, we present \name, a Transformer-based L4S queue selection system.
% Transformer architecture 
As depicted in Figure \ref{fig:overview}, \name has two components: a Transformer as a \emph{per-packet latency predictor} and a prediction-driven \emph{proactive L4S controller}. 

While L4S can benefit from per-packet queue selection, for which Transformers could be a good fit, making \name practical creates challenges in both of its components. 
% Each component presents its own challenges, and we will address them in this section.

\paragraph*{Per-packet latency prediction (\S\ref{ss:tf_predict})}
\begin{enumerate}
    \item Sharp latency changes, due to their rarity, are more difficult to predict, but they are also crucial as they are when proactive queue switching is most useful. How to make \name predict sharp latency changes accurately?
    % \emph{How to make \name predict packet latency in a way that best inform L4S queue selection?}
    % How to enable \name to focus on sharp latency changes at the packet level?}
    \item Transformers are workload-dependent, so it must be able to update itself with the new latency measurements frequently. How to fine-tune \name's Transformer with the latest measurements with minimum overhead?
    % \item \emph{How to keep \name up to date to keep making accurate predictions on changing traffic interactions?}
\end{enumerate}

\paragraph*{Queue selection controller (\S\ref{ss:tf_control})} 
\begin{enumerate}
    \item To inform per-packet marking, the latency prediction must be fast. How can \name make predictions fast enough without sacrificing its accuracy?
    \item Straightforward prediction-driven control can cause oscillation as it may assign too many packets to a seemingly low-latency queue before their ACKs are returned. How can \name prevent such oscillation? 
\end{enumerate}

\subsection{\name's latency predictor}
\label{ss:tf_predict}

\paragraph*{Transformer basics}
Transformers can learn sequentially dependent information at scale using the parallelized \emph{attention} operation. The attention module removes the need to explicitly model the sequence for learning relationships. Instead, it utilizes a weighted dot product matrix, which assigns importance to a sequence value based on its relative importance to all other values. This matrix is updated during training to understand sequential dependencies between different values. While they work well for general sequence prediction tasks, to effectively perform latency prediction, we make several custom changes to adapt to the challenges of network packet sequences and traffic patterns.

\paragraph*{Predicting sharp changes, not values} 
In network packet traces, we may have a series of packets with similar delays, interspersed with occasional sharp changes in latency.
% (\eg when congestion occurs). 
A sharp change often results from increased queuing latency as the queue might be occupied by packets from a competing flow. It is useful to select a lesser-filled queue for packets that would have otherwise experienced such an increase in queuing latency. 
Hence, we want the model to focus on learning to predict these sharp changes instead of predicting every latency value when latency changes smoothly. 

More specifically, in this paper, we define a sharp latency change as at least 20ms difference and a relative 20\% difference between two successive packet latencies.
We choose these values as in practice, a change of 20 ms and higher starts being observable for low-latency sensitive applications like VoIP~\cite{voip} and cloud gaming~\cite{cloud_gaming} on ISPs. 

\subsubsection{Focusing on sharp changes with new loss function}
\label{sss:loss-function}

% \paragraph*{Change-point focused loss function} 
\paragraph*{Why not standard loss functions?}
Standard deep learning models often use loss functions such as the Mean Squared Error (MSE) to train the models, to make the predicted value closer to the ground truth. This helps the model in learning patterns in the input data over average trends in the input. 
However, these traditional loss functions, such as the MSE, try to reduce prediction errors {\em everywhere}~\cite{zeng2023transformers}, rather than focusing on {\em sharp changes}.
% , which are more important to pinpoint. 
% Standard Transformers do not emphasise on sharp variations either, leading to less accurate predictions on these sharp changes~\cite{zeng2023transformers}. 
As we just discussed, \name needs the Transformer architecture to specifically target the neighbourhoods of sharp changes and accurately predict them for better packet-level queue selection. 
Thus, the key question is, \emph{how to make \name focus on sharp latency changes?}

\paragraph*{Focusing loss function on change points}
For this, we design a custom loss function based on the original MSE loss. We employ a scaling factor to the loss, in the case the absolute difference successive latency of two values is greater than a given threshold (\ie a sharp latency). This helps the model give greater attention to this neighbourhood, and the increased difference in latency signifies the sudden drop or spike in latency. For a given prediction $p(x)$ and a target value $t(x)$, we define the loss function $LF$ as 
\[ LF (p(x), t(x)) = 
\begin{cases}
\alpha \cdot (p(x)- t(x))^{2} \;  \text{, if} \;  | t(x) - t(x - 1) | \geq \delta \\
(p(x) - t(x))^{2} \; \text{, otherwise}
\end{cases}
\]

Our loss function $LF$ has two hyper-parameters: $\alpha$ as the penalizing factor, in case there is a sudden change in latency greater than $\delta$. The optimal choice of these values can vary with applications. 
% are workload and network dependent and can vary. 
\name's current design chooses $\alpha = 10$ and, following our definition of sharp changes (\S\ref{ss:tf_predict}), $\delta$ is at least $20$ ms difference in addition to at least a relative $20\%$ difference between two successive packet latencies. 
We observed that the \name can learn well under this design. 
% We choose these values as in practice, a change of 20 ms and higher starts being observable for low-latency sensitive applications like VoIP~\cite{voip} and cloud gaming~\cite{cloud_gaming} on ISPs. We observed that the \name can learn well under this design.  

This loss function does come with a trade-off: it has a slightly higher mean prediction error {\em averaged over all packets}, 
% performs worse on the average value prediction of \emph{all the packets} in the trace 
but is able to accurately predict many more of the sharp changes than if the model is trained with traditional MSE-based loss function. We show this in \S\ref{eval:ss3}. For our L4S queue-selection, we need \name's predictor to accurately predict latency spikes and drops for packet-level queue selection.

\subsubsection{Fast online model fine-tuning}

% \paragraph*{Online model fine-tuning} 
We train the \name to learn the interaction between packets from their implicitly embeddings in the packet sequence. These interactions can significantly change with changing traffic patterns. For example, if new flows start, flow interactions can significantly change.  
Adapting to such \emph{concept drift} is necessary for a model like \name.

The Transformer architecture is known for its fine-tuning efficiency compared to other neural network models. 
It can perform comparatively fast model updates, but the fine-tuning speed of traditional Transformers (\eg used for language modeling) is still too slow for packet-switched networks. 
Commonly used Transformers take tens to hundreds of minutes to fine-tune~\cite{ft_time}, which is too slow to keep up with changing network traffic. 
We observed that to use \name's predictions on long traces with new flows, fine-tuning the model is needed for continued correct sharp change predictions.

To further speed up fine-tuning, we design the Transformer architecture to have \emph{fewer} trainable parameters with two Transformer Encoder-based \emph{attention block} layers followed by two \emph{linear layers} derived from the PyTorch~\cite{pytorch} implementation. 
% This is followed by two \emph{linear layers} of size $d_{model} = 64$ as shown in Figure~\ref{fig:architecture}. 
In total, it has $\sim100K$ trainable parameters. This is much smaller than most Transformer models over the years, which have parameters in the order of $100M-400B$~\cite{bert, llama2, llama3}. 
To further speed up the fine-tuning, we update {\em only} the linear layers (includes activation) of the model as \name has already learnt the representation of latency changes and just needs to focus on new interactions in the recent network state. 

% We perform fine-tuning at fixed intervals, which is chosen as a value for which prediction quality does not significantly drop. 
To minimize the compute overhead of fine-tuning, we do not want to fine-tune excessively with marginal gains as we want to prevent resource wastage.
% (we currently use a one GPU). 
To find a reasonable fine-tuning interval, we independently measured the performance at different fine-tuning intervals, starting from the same pre-trained model. We found a common regular interval which performs well across our workloads. We evaluate this in \S\ref{eval:ss3}. 

\begin{figure}[t]
    \centering
    \includegraphics[width=0.82\linewidth, keepaspectratio]{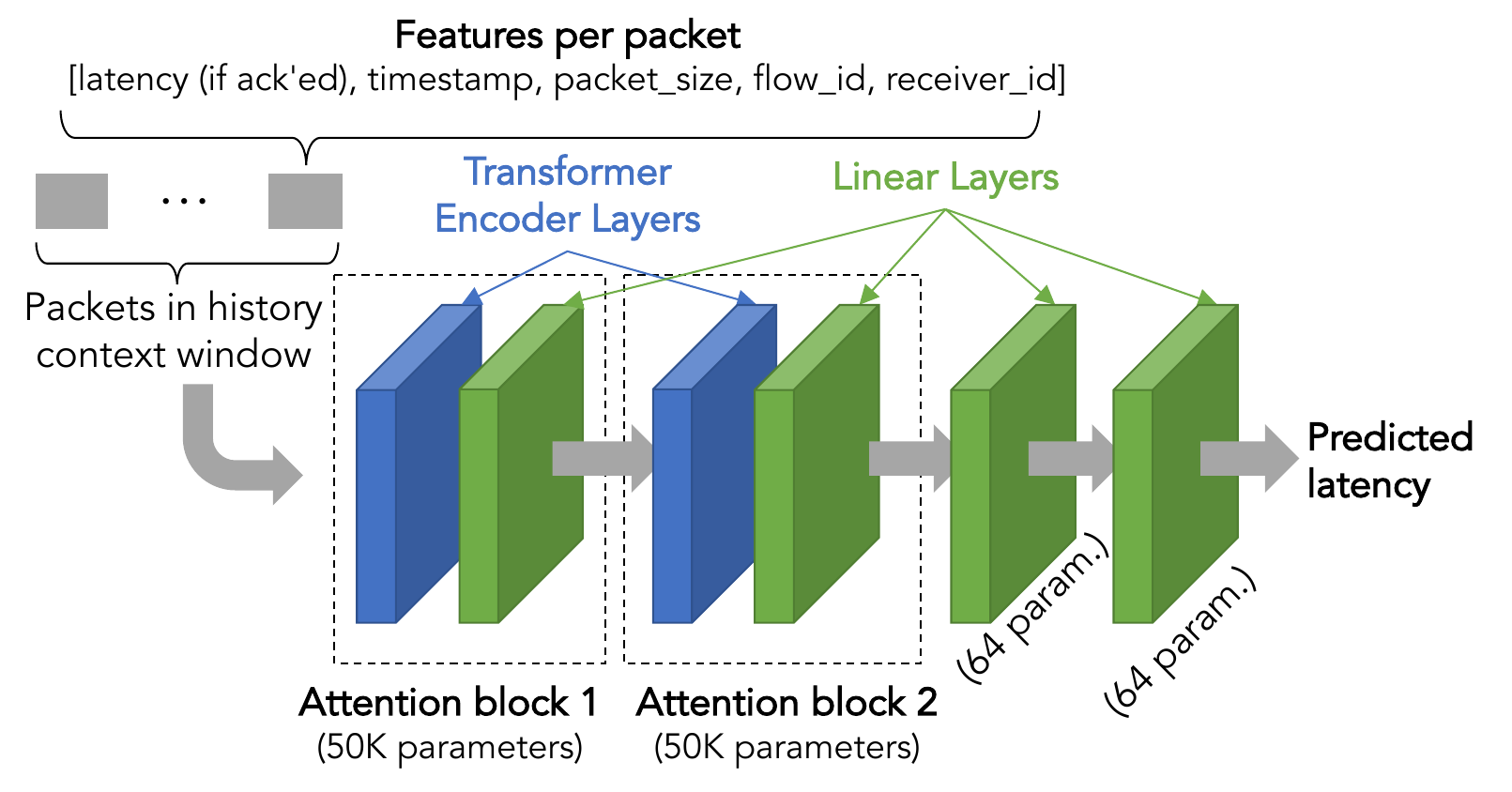}
    \caption{\name's architecture with showing its context window, multiple \emph{attention blocks} and \emph{linear layers}.}
    \label{fig:architecture}
\end{figure}

\subsubsection{Input features of the latency predictor}

In addition to our customization of the loss function and the model architecture, we also take care in designing the {\em input} of \name's latency predictor.
As part of the Transformer's input, the end-to-end latency of each acknowledge packet includes the queuing delay of the packet and the propagation delay on the link. Thus, the latency value is influenced by the interaction between packets on the network, and it reflects the effect of events queue buildup, congestion, and packet loss. 

That said, the past latency values {\em alone} are insufficient to 
% ensure enough information is captured to 
% consider two more design components to 
make \name learn effectively. 

\paragraph*{Input features beyond just latency values} 
The parallelized attention mechanism lose the explicit {\em position} of each item in the sequence and must be supplied with this information externally in the form of \emph{positional embeddings}. Positional embeddings can take many forms~\cite{shaw-etal-2018-self, vaswani2017attention, Gehring2017ConvolutionalST}. 
Fortuitously, network traces \emph{naturally} provide the positional information with \emph{packet timestamps}. 
We use the \emph{relative timestamp} of a packet with respect to the first packet of the flow as our absolute positional information. 

Moreover, we discovered that only using past latency values and timestamps does not provide enough context about the network for learning latency prediction. 
We select a few more features in order to indicate the state of the network. 
Inspired by the authors in~\cite{10.1145/3563766.3564104} where they show the need to jointly look at both packet-level and network-level features, for \name we choose the following features as input to supplement packet latency values and timestamps. 
% packet\_size, end\_to\_end\_latency, flow\_id and receiver\_id. 

% \begin{figure}[h]
%     \centering
%     \includegraphics[scale=0.5, keepaspectratio]{figs/attn_heatmap_agg_0_2.pdf}
%     \caption{Attention weights visualization with \name on a NS3 simulated trace of a 200 packet sequence measured from a portion of 60 concurrent TCP Cubic flows}
%     \label{fig:attention}
% \end{figure}

\begin{itemize}
    \item \emph{packet\_size:} The packet size along with the timestamp implicitly helps \name understand the real-time sending rate. The size of the packet affects the number of packets which can be accommodated in the queue buffer, \eg larger packets means fewer packets are present in the queue. Using the packet size helps \name understand packet interactions inside the queue better.
    \item \emph{flow\_id:} Assigned to all packets of the same flow with the same value, the flow\_id lets \name distinguish between the packets from different concurrent flows from the same sender.\footnote{\name's current design requires the visibility of the latency and L4S decision on all packets being sent out from the same end-point in order to make accurate predictions. 
    Note that this does not require visibility into each flow's internal TCP state machine.
    % This requires an abstracted view of all individual TCP state machines across every flow. 
    % Implementing this in the Linux kernel requires a new module, which can jointly view states from multiple TCP streams. 
    Implementing this in the Linux kernel requires a new module that can access multiple TCP streams. 
    While we do not focus on the exact implementation of such a module in this paper, several recent works~\cite{ccp_1, ccp_2, ccp_3} have developed a centralized TCP CC plane, which views multiple data paths and integrates a central feedback loop. Several of these already have Linux kernel integration. Compared to these efforts, our design has a much simpler requirement---it only needs to look at flows from the same endpoint. Thus, we believe a portion of an existing centralized TCP CC module can be reused by us for \name.}
    % , of which a possible implementation will be outlined in \S\ref{??}. 
    This helps \name understand interactions of packets between flows better.
    \item \emph{receiver\_id:} Assigned to all packets being sent to the same receiver endpoint, the receiver\_id helps \name understand which packets likely traverse the same path and the effect of different path latencies.
\end{itemize}

It is true that adding more features may further improve learning, but these features are already sufficient for \name to learn to predict latency at the sender. 
% we identified the above features as essential for learning to predict at the sender.

% To visualize the \name's attention applied on dependencies between packets, in Figure \ref{fig:attention} we plot one layer's attention weights (averaged over all 8 attention heads in \name's encoder) of a trained Transformer for a 200 packet sequence. Each packet is a combination of the above mentioned features. At each layer the Transformer learns mappings between different positions in the packet sequence, reflecting in the attention weight values.

% \paragraph*{Transformer scaling and history size}
\paragraph*{History context window size}
General Transformers scale quadratically $O(n^{2})$ with input sequence size $n$ (in tokens). 
However, sequences in network traces are often over $1000s$ of packets in length for a trace measured over a few seconds. Since learning from extremely long sequences can scale poorly, we need to choose 
% how far into the past we look. 
\emph{how many packets into the past do we have to look into to learn sharp latency changes?} 

Since sharp latency changes in packet sequences are often constrained  to localised neighbourhoods of effect (\eg a queue fills up and drops packets), we observed in practice we can use shorter histories of packet sequences to train \name's predictor. 
In particular, we use a sliding window approach, where we train the Transformer on smaller sliding windows of history size rather than the entire trace, reducing the quadratic scale factor. 

Rather than choosing a fixed window size, \name matches the sliding window size to the maximum \emph{bandwidth-delay product} (BDP) of the link. 
This roughly covers the number of in-flight packets in the network at all times and we observed that packets which are queued together have greater interaction over the current network state. For example, a recent packet history of $1-2\times$BDP can signal queue buildup. We can dynamically adapt the sliding window during training to match changing BDP values for various network conditions. We allow \name to learn these interactions over multiple sliding windows. Of course, our choice of history window size may not be optimal, and more research will be needed to pick the optimal window size. 
% We do not claim this is the universally optimal value for the history sequence and acknowledge for many latency change events, we might have to much further into the past. We leave further investigation on that to future work.

\subsection{\hspace{-0.cm}\name's L4S queue selection}
\label{ss:tf_control}

We design \name to predict latency and act as a packet-level controller for queue selection at the sender. After \name predicts whether a set of subsequent packets is likely to experience high latency, we leverage the L4S's packet header marking to influence the queue selection at L4S dual-queue routers.
% we need to guarantee that \emph{packets are sent to only to the chosen queue}. We 
% we leverage the L4S strategy where the router's naturally separate dual-queue setup has specific conditions on packet marking and queue selection using the ECN bits in the IP header. 
We explain this in more detail next.

\paragraph*{L4S queue selection with \name}
Current L4S queue selection involves putting all packets from a given L4S flow into the same queue. 
Yet, recall that this approach can be problematic when dealing with localised increasing cross-traffic congestion in a short window (\S\ref{subsec:queue_selection}). Such events result in higher packet latency, pushing up the tail latency. % for the entire flow. 
Predicting packet-level latency at the sender enables marking packets within a flow to be put into different queues to reduce latency.

In practice, L4S routers are designed to respect ECN markings. Thy are configured to handle both classic flows (i.e., non-L4S flows with packets are marked \emph{non-ECT}) and L4S flows (packets marked as \emph{ECT}). Once packets of a flow are marked, every L4S-enabled router downstream respects the packet marking. L4S traffic is sent using DCTCP based TCP Prague, with all packets are marked as ECT. L4S enabled routers at every hop, ensure that ECT marked traffic is placed on the L4S queue and all other traffic is placed in the classic queue. If a router does not have L4S support all traffic is placed in the classic queue.

For a set of L4S flows from a sender, \name predicts sharp latency spikes for future packets based on the learnt effects of packet interaction. 
% As L4S allows natural queue separation, 
Based on the prediction, \name's controller makes a simple decision in order to reduce the queuing latency on the L4S queue on the bottleneck router. It changes the individual packets within the flow which have high predicted latency to have non-ECN markings in the IP header. These packets will then be put on the less-congested classic queue on the bottleneck router. 
% We illustrate this in Figure \ref{fig:overview}, where the predicted latency is used to select the queue at the packet-level. 
This reduces the queuing latency experienced by these L4s flows.

% It is important to note that the \name can achieve significant latency reduction during the times that the classic queue is relatively less congested. 
We acknowledge that if {\em both} queues are at capacity, changing the queue will not reduce latency. For this paper, we focus on reducing the tail latency for L4S flows, when the classic queue is relatively less congested. We also do not change the router behavior or modify the sending congestion control (CC) logic. Instead, we only perform queue selection. As a result, \name's reduction of tail latency should have little negative effect on throughput.
% This prevents \name's controller from reducing the throughput.

\subsubsection{Fast queue selection for in-time predictions}

% \paragraph*{Making prediction and queue selection fast enough} 
In order to use the \name's predictions for L4S queue selection, the predictions need to be \emph{fast enough}. This means we need to make the decision before it is time to send out the packet. High link speeds and data rates require an extremely fast inference (for both prediction and control logic). Waiting for the last packet to make a prediction is not always scalable at the packet level with increasing link speeds.

\begin{figure}[t]
    \centering
    \includegraphics[width=0.9\textwidth]{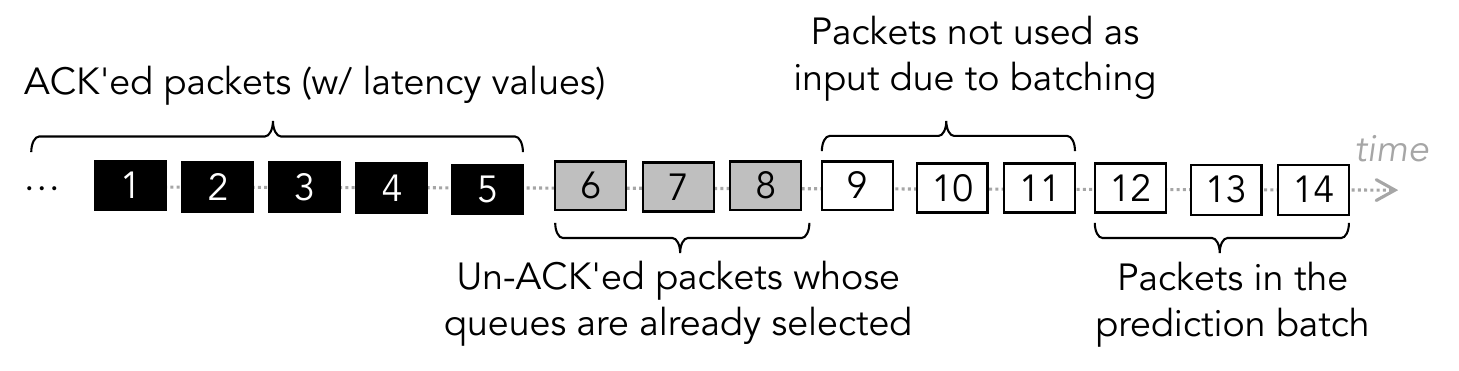}
    \caption{An illustration of two main ideas.
    (1) \name speeds up inference by batching, \eg Prediction of Packet 12-14 is done simultaneously as a batch, though the prediction delay precludes the use of the information between Packet 9-11.
    (2) \name prevents prediction-driven oscillation by incorporating un-ACK'ed packets in the prediction input, \eg although Packet 6-8 do not have latency values yet, but by incorporating their queue selection in the input, the predictor knows how many packets have been assigned to each queue.} %\name's design for fast inference and queue selection if packets are unACKed}
    \label{fig:batching}
\end{figure}

To solve this, we make \emph{batched} predictions. Like in standard Transformers, batching increases the throughput with a negligible increase in inference delay because running the model on one input does not fully utilize the compute resource. 
We make latency predictions for a future batch of packets, given the latency values of the preceding packets.  By increasing the batch size for predictions, the inference latency at the per-packet level is reduced. Since the prediction for a future packet is made a few packets earlier, \name's controller has time to account for the inference delay and the control logic, in order to make a decision for the packet before it is sent out. 
Figure~\ref{fig:batching} shows an illustrative example. 
We need to make the prediction for \emph{packet 12-14}, as one prediction batch, by seeing up-to \emph{packet 8} in order to ensure we send out \emph{packet 12} in time. 

However increasing the batch size beyond a point does come at the cost of reduced prediction quality. \name's current implementation finds a balance point where it supports a large enough prediction batch for in-time predictions on fast ISP links, without significant drop in quality.  Currently \name's predictor is fast enough to make predictions for real networks like ISPs etc. running its typical applications (e.g. YouTube video streaming at 1080p).
We can predict upto a batch size of packets (upto 8 future packets) within $200-400$ $\mu$s of inference latency without noticeable reduction prediction quality. This is fast enough for predictions and decisions in time for link speeds of upto 200 Mbps.

\subsubsection{Preventing oscillation in selected queues}

% \paragraph*{Preventing oscillation}
% \paragraph*q{What if we do not have the latency of the previous packets in time} 
Finally, pure prediction-driven control often leads to unstable outcomes. 
It is possible that \name's queue selection controller can incorrectly select a particular queue for a packet if it only uses information from packets which have been ACKed. If \name has not received enough recent acknowledgments (ACKs) for the previously sent out packets, it may not select the queue correctly as it does not have the most recent network state information. This can cause \name to overfill a particular queue and increase the latency.
% . Being inaccurate in queue selection for these particular packets can lead to increased latency. 

To prevent this, \name's queue selection controller additionally uses a combination of 
(1) latency values of packets that have been ACK'ed and (2) packets that have been marked with queue-selection decisions but have not been ACK'ed. 
% available previous packet latency values (for which ACKs have been received) along with viewing the queue selection for the packets sent out on the port, which have not been ACKed yet. 
A combined view of these factors helps \name decide if the packet should be finally put on a particular queue in case the previous latency values are missing. 
We illustrate this in Figure~\ref{fig:batching}. 
For instance, \name's queue selection for \emph{packet 12} looks at both past packets with latency values (\emph{packet 1} - \emph{packet 5}) {\em and} at the queue selection on un-ACK'ed packets (\emph{packet 6} - \emph{packet 8}) in order to select the queue for the given packet.
Even though we do not know the latency of the un-ACK'ed packets, they still provide vital information about at least how many packets have been assigned to each queue. This information is helpful for \name to prevent oscillation in its queue selection decisions.

\section{Implementation}
\label{sec:implementation}

We implemented \name's predictor and L4S queue selector in 1500 lines of Python code using PyTorch~\cite{pytorch} and Lightning~\cite{Falcon_PyTorch_Lightning_2019}. We build on an existing NS-3 official implementation of L4S~\cite{ns3_l4s}. The L4S applications within NS-3 were implemented in 500 lines C++ code.

For training and inference we use a single TITAN RTX 3080 with 10 GB memory with a peak memory utilization of 2GB. This is {\em much cheaper and less powerful} compared to GPUs used to train Transformers today. 
For comparison, LLMs today use multiple H100 GPUs for training, and even a single H100 GPU has $10-12\times$ higher memory and has $>100\times$ higher price than what we use to run \name.

As we train for prediction tasks, we only use the Transformer Encoder layers to learn representation similar to the literature on Transformers for time-series as introduced in Section \ref{subsec:existing_work}. We use the ADAM~\cite{adam} optimizer with its default parameters of $\beta_{1} = 0.9 $, $\beta_{2} = 0.98$ and $\epsilon = 10^{-9}$.

We use dropout rate of $0.2$ for both the Transformer and the linear layers and ReLU~\cite{relu} activation for non-linearity. We also employ a weight delay of $1\epsilon^{-5}$ to help with regularization and reduce over-fitting. We vary the learning rate based on the original Transformer paper~\cite{vaswani2017attention} as $ l\_rate = d_{model}^{-0.5} \cdot \min (step\_num^{-0.5}, step\_num \cdot warmup\_steps^{-1.5})$ with setting $warmup\_steps = 2000$, based on empirical testing.

% All models are trained, fine-tuned and evaluated exactly the same way.

We have an initial pre-training time of 5 - 10 minutes on the real traces collected in Table \ref{eval:table0}. We do not optimize the pre-training time as it is a one-time cost. We instead focus on fine-tuning and inference speeds, as they are of greater, practical significance since the models need to be frequently updated (more details in \S\ref{sec:eval}). 

Finally, we use {\bf \em separate} data for training, fine-tuning, and testing, \ie no data pollution, and they are {\bf \em aligned} across \name and its baselines (\S\ref{sec:eval}).
We first pre-train \name and the baselines on the first 5 minutes of the traces. We then test them on the next 5 minutes of the trace. At every 5 minute intervals, we fine-tune the models on the preceding 1 minute of the trace, and then again test on the next 5 minutes till the end of the trace.

\begin{table*}[t]
    \centering
    \renewcommand{\arraystretch}{1.35} % increase row height by 30%
    \scalebox{0.75}{
    \begin{tabular}{ l  c   c   c   c   c}
    \toprule
         Trace Name &  Description & Median Latency & Trace Duration & Flow Duration  & \# flows \\
            &  &    (ms) & (min) & (s) & in trace\\
         \midrule
         {\em video\_wifi} & Netflix traffic captured on a WiFi endpoint & 70 & 30 & 29 - 386  & 28 \\
         {\em web\_wifi} & Facebook traffic captured on a WiFi endpoint & 66 & 30 & 27 - 172  & 42 \\
         {\em video\_eth} & Netflix traffic captured on an Ethernet endpoint & 74 & 30 & 48 - 362 & 31 \\
         {\em web\_eth} & Facebook traffic captured on an Ethernet endpoint & 57 & 30 & 35 - 212 & 47\\        
    \bottomrule
    \end{tabular}
    }
    \caption{Summary of the collected traces on the WiFi and Ethernet endpoints}
    \label{eval:table0}
\end{table*}

% \begin{table*}[t]
%     \centering
%     \begin{tabular}{ l  c   c   c   c}
%     \toprule
%          Trace Source & Median & P90 & P99  & \# flows\\
%             &    (ms) & (ms) & (ms) \\
%          \midrule
%          Netflix (WiFi) & 70 & 200 & 550 & 68\\
%          Facebook (WiFi) & 66 & 160 & 340 & 52\\
%          Netflix (Ethernet) & 74 & 190 & 480 & 66\\
%          Facebook (Ethernet) & 57 & 140 & 290 & 63\\        
%     \bottomrule
%     \end{tabular}
%     \caption{Latency statistics on the collected traces on WiFi and Ethernet \sr{update}}
%     \label{eval:table0}
% \end{table*}

\section{Evaluation}
\label{sec:eval}

The key takeaways from our evaluation are 

\begin{itemize}
    \item \name predicts \emph{sharp changes} in latency more accurately than other machine-learning baselines by 45-65\% across a variety of real network and simulated traces.
    \item \name achieves lower P99 tail latency on L4S flows from a given sender by 36-45\% compared to the default L4S queue selection logic as well as queue selected by other latency predictors. % using its packet-level queue selection
    \item \name allows frequent fine-tuning to keep the model up to date over changing network conditions and makes latency prediction fast enough at a line speed of 200Mbps.
\end{itemize}

\paragraph*{Baselines}
We choose the following baselines as they were \emph{state-of-the-art} for time-series prediction tasks prior to the introduction of Transformers.
\begin{itemize}
    \item {\bf LSTM:} We use a state-of-the-art BiLSTM~\cite{bilstm} model known used in the past for its sequence prediction capabilities. For fairness, we align the LSTM model to have comparable size (parameter count) as \name (albeit having much higher inference and fine-tuning times).

    \item {\bf XGBoost:} We use a state-of-the-art classical ML regression model XGBoost(XGB)~\cite{xgboost}, which has been been used in the past for certain sequence prediction tasks using sliding windows of input sequences~\cite{xgboost_pred}.
\end{itemize}

\paragraph*{Evaluation Metrics} We evaluate the prediction of \name and the baselines on the packets where there is a sharp change in latency. We define a \emph{sharp change} as a change in latency between two consecutive packets of \emph{at least 20 milliseconds} and a \emph{relative change of at least $20\%$} as introduced in \S\ref{ss:tf_predict}.  We use \emph{precision}, \emph{recall} and \emph{F1-score} to measure prediction quality at these latency changes. We define success if both the true latency and the model's predicted latency have a \emph{sharp change} in the same direction. We report the prediction quality only over the packets that have these sharp changes in latency. For the tail-latency improvement, we measure the $\%$ reduction in P99 queuing latency. We refer to this metric from now on as \emph{P99 Reduction}.

\subsection{Evaluating \name's predictions and L4S queue selection on real network traces}
\label{eval:ss1}

% We evaluate \name's predictions on a varying set of real network traces. Traces are from real measurements and are not replayed using collected statistics in an emulator.

% \begin{itemize}
%     \item Home network (Service recognition dataset, Netflix flows)
%     \item Campus network (VCA dataset?)
%     \item More?
% \end{itemize}

\begin{figure*}[t]
    \centering
    \includegraphics[width=\linewidth, keepaspectratio]{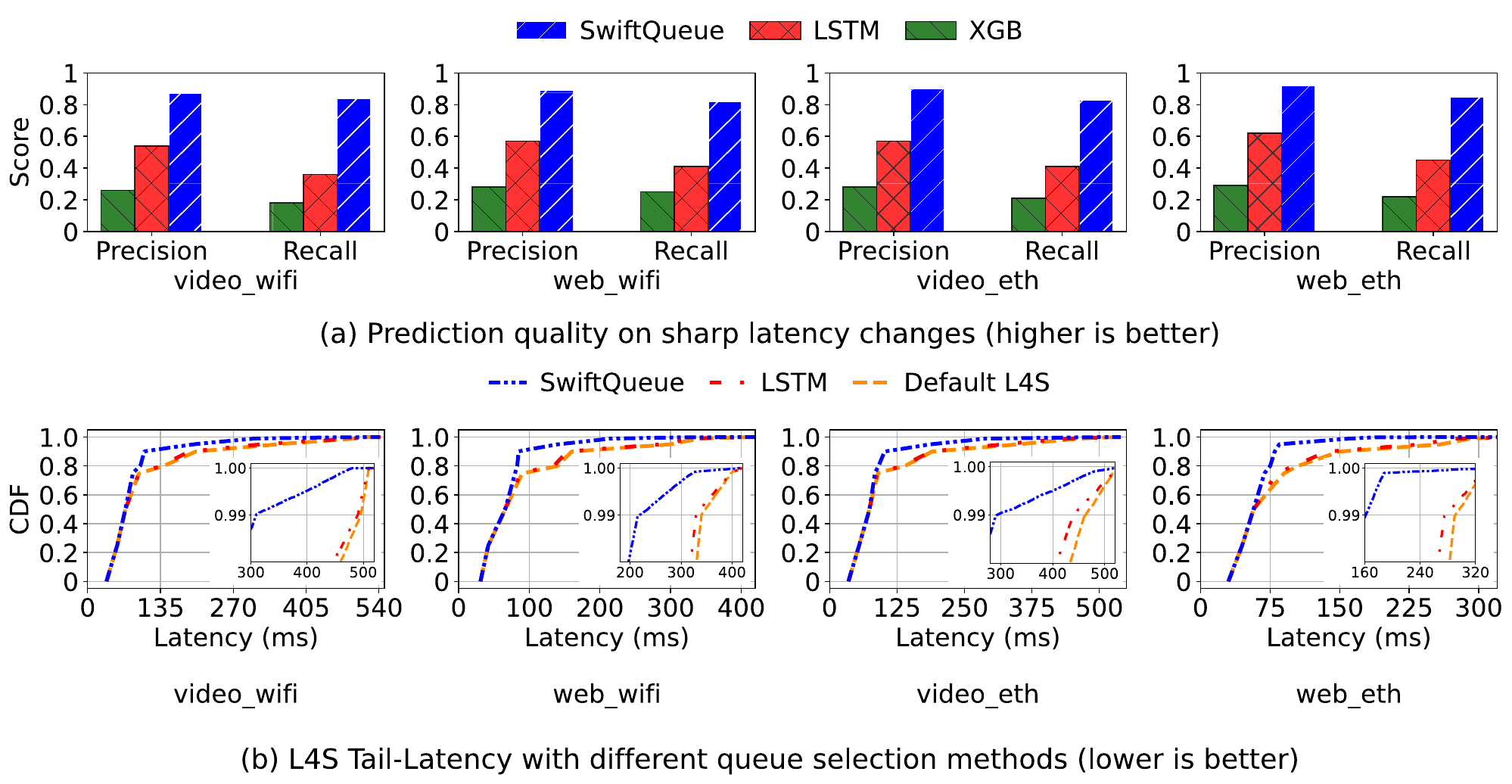}
    \caption{\name achieves better prediction on sharp changes by 45-65\% and has higher tail-latency (P99) reduction for L4S flows by 36-45\%, compared to other predictor-driven queue selection or L4S's default strategy.}
    \label{fig:eval_real_clf}
\end{figure*}

\paragraph*{Real-world traces collection} We capture Netflix traffic (video streaming application) on a Home-WiFi network and on a wired Ethernet network. We also collect Facebook browsing traffic (social media web traffic application) on the same Home-WiFi network and Ethernet network. All of the flows captured are from within the same browsing session and we define a flow as the standard 5 tuple. We use self-collected ISP traces as popular public traces like CAIDA~\cite{caida} or M-LAB~\cite{mlab} may not have all the features required for us.
% \footnote{If accepted, we will make the traces public for further community research} 
All traffic is collected on the same end-point device and the available bandwidths on the WiFi and Ethernet links were measured to vary between 50-100 Mbps. Table \ref{eval:table0} shows the details of the collected traces.

\paragraph*{L4S queue selection} We use trace-driven simulation in NS-3 to evaluate the performance of \name's L4S queue selection. We use statistics such as trace duration, inter-packet gaps, packet sizes etc. from the real traces in Table \ref{eval:table0}.

\begin{figure}[t]
    \centering
    \includegraphics[width=0.7\textwidth]{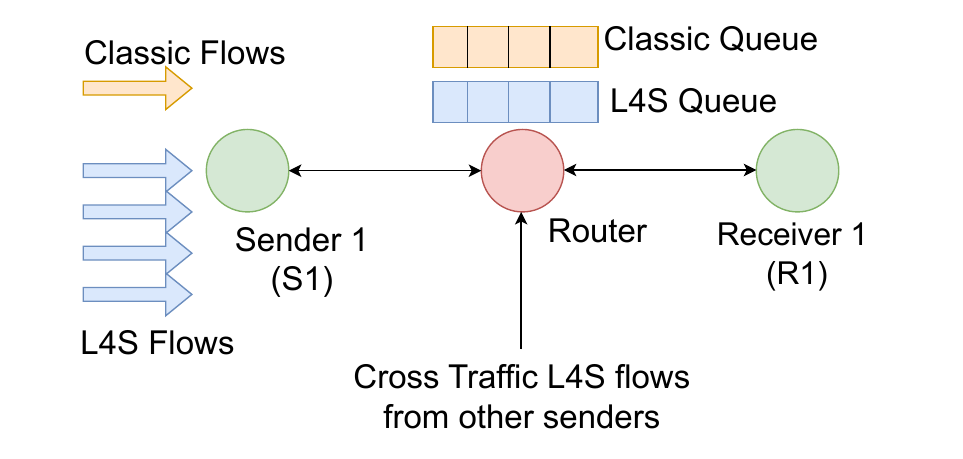}
    \caption{L4S queue selection topology in NS-3 with both L4S and Classic flows.}
    \label{fig:l4s_setup}
\end{figure}

 We use an L4S implementation in NS3 with a dual-queue setup as specified by the authors in \cite{l4s}. We setup a L4S enabled topology as shown in Figure \ref{fig:l4s_setup}. We have a \emph{6 L4S flows} along with \emph{4 Classic TCP Cubic flows} bound from Sender 1(S1) to destination Receiver 1(R1) . We have \emph{10} L4S flows from different senders creating cross traffic (unseen by \name at S1). All L4S flows share the same bottleneck L4S queue on the router.

By default, all packets from all L4S flows will be put on the L4S queue on Router (Default L4S). We use \name's predictor to predict the latency of next packets at \emph{S1} and change the ECN marking for the packets with high latency. These L4S flow packets from sender \emph{S1} are put on the classic queue. As there is a large number of L4S flows, they experience increased queuing latency on the L4S queue. We \emph{focus} on reducing the tail latency for the L4S flows sent out by Sender \emph{S1}. We reduce the tail latency of packets for the L4S flows bound for R1 by proactively changing the queue selection to reduce the queuing latency.

% \begin{figure*}[t]
%     \centering
%     \includegraphics[width=0.74\linewidth, keepaspectratio]{figs/sensitivity_analysis.pdf}
%     \caption{\name has better prediction on \emph{sharp changes} versus baselines over changing queue sizes and \# of flows (NS3-simulated traffic)}
%     \label{fig:sensitivity}
% \end{figure*}

For measuring the improvement of packet-level queue selection, we report the \emph{P90 and P99 latency} for the L4S flows. We compare with the default L4S selection (and not other AQM techniques) as we don't change the L4S logic within the router. We only change the selection logic at the sender and do not adapt the congestion control end-point to modify its sending rate.

\textbf{Result \#1 - \name can predict sharp latency changes better compared to the baselines} 
Figure \ref{fig:eval_real_clf} (a) shows the performance of the \name relative to the LSTM and XGB baselines in predicting the sharp end-to-end-latency changes for the next packet. 
\name predicts the sharp changes in latency much  better than the LSTM and XGB baselines, consistently being \emph{45-65\%} higher values times better for all the real-world traces collected, in terms of both precision and recall. 

\textbf{Result \#2 - \name's prediction driven L4S queue selection achieves lower tail latency than the baselines} Figure \ref{fig:eval_real_clf} (b) shows the improved tail-latency reduction of the L4S flows by \name's prediction based queue selection. \name's queue selection achieves \emph{36-45\%} P99 Reduction over the LSTM baseline and the default L4S selection.

\subsection{Evaluating \name's predictions on NS-3 simulated data}
\label{eval:ss2}

\begin{figure}[t]
    \centering
    \includegraphics[width=.75\linewidth]{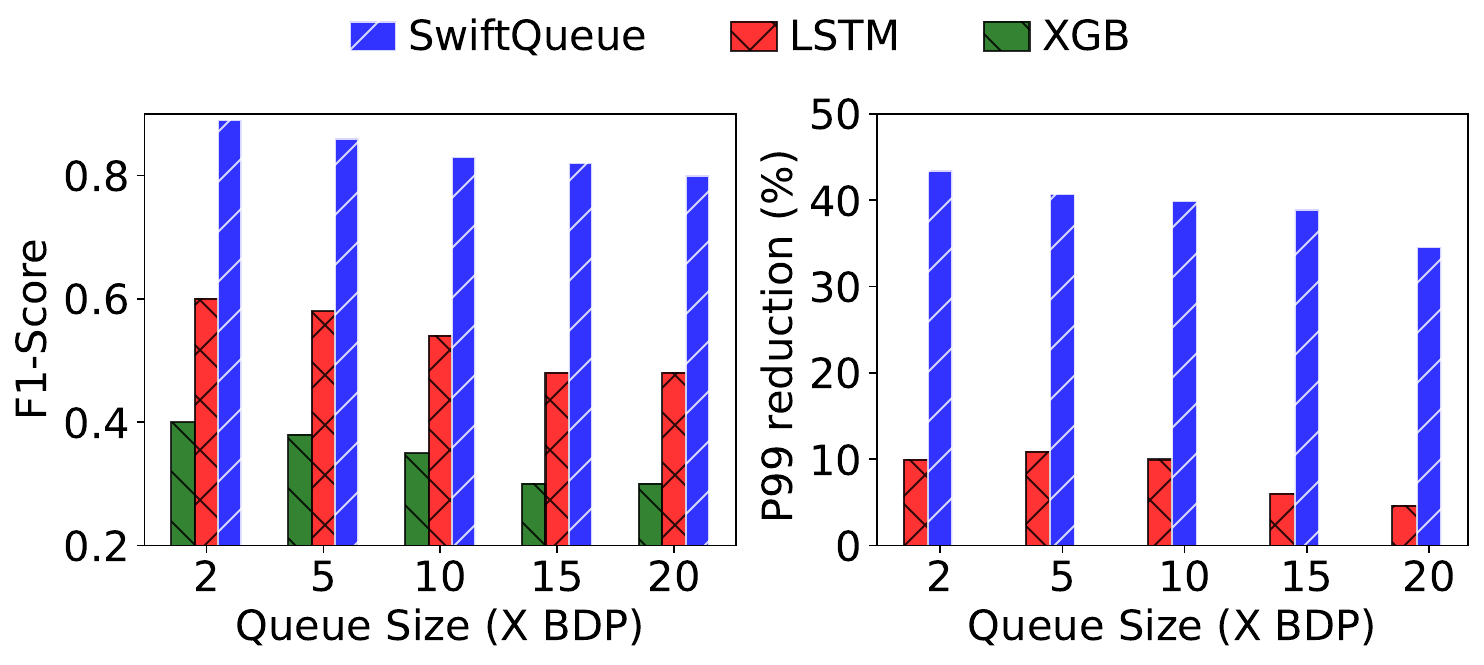}
    \caption{Impact of the bottleneck queue size on prediction accuracy and tail latency reduction.}
    \label{fig:sens_queue_size}
\end{figure}

\begin{figure}[t]
    \centering
    \includegraphics[width=.75\linewidth]{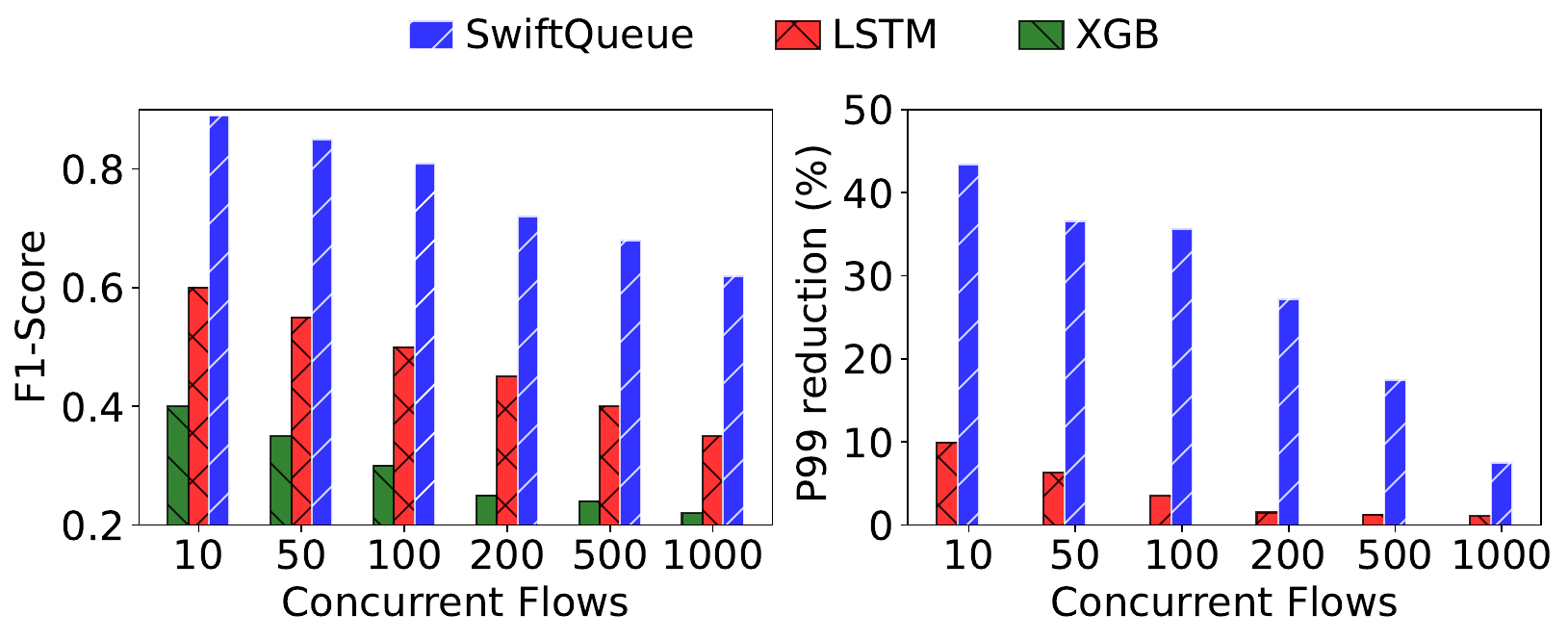}
    \caption{Impact of the number of flows on prediction accuracy and tail latency reduction.}
    \label{fig:sens_num_flows}
\end{figure}

\paragraph*{Evaluation Setup} To further evaluate the performance of \name's predictions on packet-level interactions, we use NS-3~\cite{ns3} to simulate workloads for training and evaluating on varying network conditions. We reuse the three-node topology (one sender, router and receiver) as shown in Figure \ref{fig:l4s_setup} for our experiments. We only use TCP flows for our evaluation as we need a reliable, connection-oriented protocol. While the topology is simple, we create complex packet and flow interactions.

The TCP flows are created using messages sampled from real world distributions~\cite{10.1145/3230543.3230564}. We use the web search workload for DCTCP~\cite{10.1145/1851182.1851192} and the Hadoop cluster workload at Facebook~\cite{10.1145/2829988.2787472}. Our default environment runs 10 concurrent TCP flows using DCTCP+L4S and 4 additional classic TCPCubic flows (for cross traffic). This is a realistic setup, as in practice L4S flows do not usually exist without other Classic flows on the network. We use a default propagation delay of $20 ms$ and a link bandwidth of \emph{100Mbps}, reflecting a typical ISP connection. Each flow is started with a base sending rate \emph{5Mbps}. We use a default queue length of $2\times$BDP. To ensure we have enough packet interactions, flows are started in a flow start window, according to a uniform random distribution. This ensures the flows start at different times leading to greater interaction between flow start times.

We perform a sensitivity analysis to measure the performance of \name in terms of prediction quality and P99 latency reduction for the L4S flows. We evaluate along the following dimensions. We change one dimension at a time, keeping all others the same as our default environment.

\begin{itemize}
\item {\bf Queue size:} We increase the queue size on the bottleneck router as 2, 5, 10, 15, and 20 $\times$ the \emph{BDP} respectively.
\item{\bf \# of concurrent flows:} We increase the number of TCP flows using DCTCP+L4S as 10, 50, 100, 200, 500 and 1000.
\item {\bf Propagation delay:} We increase the propagation delay of the link as 20, 40, 100, 150 and 200.
\item{\bf CC algorithm:} We change our 10 concurrent TCP flows to run DCTPC+L4S, TCPCubic and TCPBBR. We choose them as they represent loss-based~\cite{tcp_cubic} and delay-based~\cite{tcp_bbr} CC algorithms that are deployed today.

We compare \name's performance against the baselines on 5 minutes of simulated traffic. We don't use the XGB model to perform packet level queue selection as the prediction quality on the \emph{sharp changes} is extremely low.

% \begin{figure}[h]
%     \centering
%     \includegraphics[width=0.47\textwidth, keepaspectratio]{figs/l4s_ns3_latency_envs.pdf}
%     \caption{\name achieves greater reduction in P90 and P99 latency compared to the baselines across the NS-3 experiments}
%     \label{fig:ns3_l4s_latency}
% \end{figure}

\paragraph*{Result  - \name achieves better predictions changing new network conditions and obtains lower P99 latency on the L4S flows} We compare the prediction quality of \name with that of the baselines over all the individual environments. We use F1-score on the \emph{sharp changes} as a measure of the prediction quality. We use the $\%$ decrease in P99 latency on the L4S flows, relative to the default setting of queue selection at the flow level in order to compare the performance of \name's queue selection controller.

\begin{figure}[t]
    \centering
    \includegraphics[width=.78\linewidth]{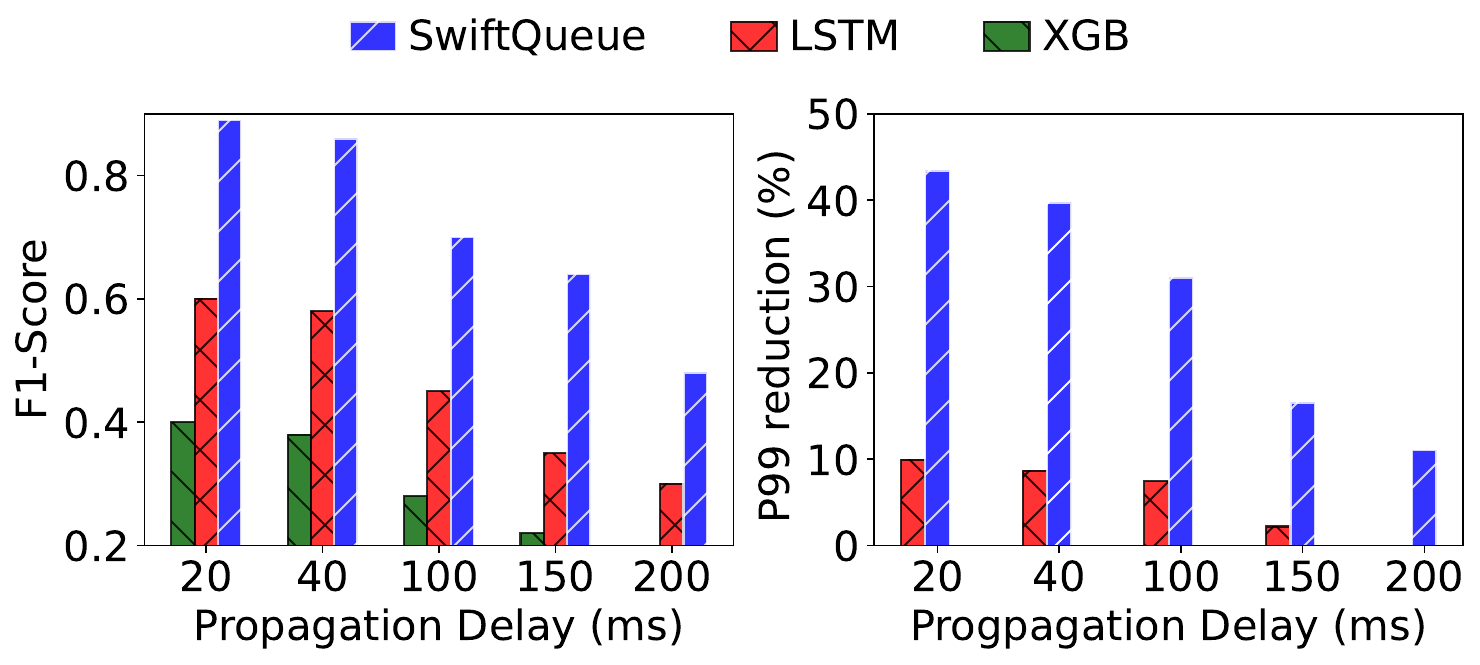}
    \caption{Impact of the end-to-end propagation delay on prediction accuracy and tail latency reduction.}
    \label{fig:sens_ld}
\end{figure}

\begin{figure}[t]
    \centering
    \includegraphics[width=.8\linewidth]{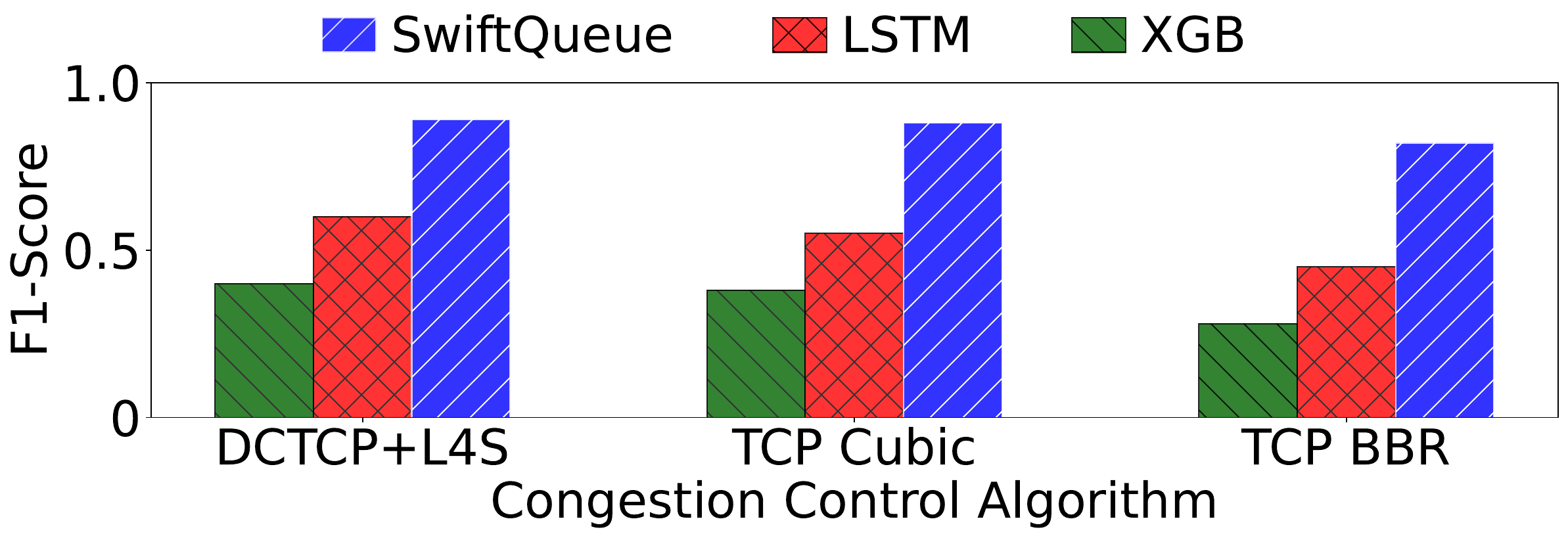}
    \vspace{0.1cm}
    \caption{\name's latency prediction accuracy under different congestion control logics.}
    \label{fig:sens_cc}
\end{figure}

First, in Figure \ref{fig:sens_queue_size}, we measure the effect of changing the max queue. \name performs better than both the baselines in terms of higher F1-score by \emph{29-62\%} across all the environments and also achieves a greater P99 Reduction by \emph{33-35\%} across all environments. \name is not sensitive to increasing queue size for a given number of concurrent flows and can continue to perform well for both prediction quality and latency reduction. With a given number of flows, changing queue size doesn't increase the complexity of packet interactions over time due to fewer, new flow starts.

Second, in Figure \ref{fig:sens_num_flows}, we measure the effect of changing the number of concurrent flows on the performance of \name. \name  performs better than both the baselines in terms of higher F1-score of \emph{29-60\%} across all the environments. \name also achieves a greater P99 Reduction by \emph{10-35\%} across all environments. We expect the performance of \name to be slightly sensitive, with increasing number of flows as interactions can get extremely complex. However, even upto 100s of concurrent flows, \name provides useful latency prediction and P99 latency reduction.

Third, in Figure \ref{fig:sens_ld}, we measure the effect of changing the link's propagation delay on the performance of \name. \name performs better than both the baselines in terms of higher F1-score by \emph{29-65\%} across all the environments and achieves a greater P99 Reduction by \emph{} across all environments. \name is sensitive to increasing propagation delay on the link beyond a point for a given number of concurrent flows. With increasing propagation delays, the ACKs received at the sender are often delayed and in such cases, \name may not have enough past context to predict accurately or perform queue selection. However, even at high propagation delays of $200$ ms, \name still has observable gains and does not drop quality or increase latency.

Finally, in Figure \ref{fig:sens_cc}, we measure the effect of only using a single TCP CC algorithm. We see that across several popular TCP CC algorithms, \name has better performance in terms of prediction F1-score by \emph{25-49\%}. We observed that TCP BBR flows have slightly more random interactions and the prediction F1-score is slightly lower (about 4-5\%). We feel this happens due to the probing nature of TCPBBR, which sends out special probing packets. \name currently cannot distinguish between such control packets explicitly and we understand this might be a potential reason for the reduced quality. Finally, we don't evaluate the P99 latency reduction for TCPCubic and TCPBBR as they cannot be configured to enable packet-level queue selection using ECN markings

\end{itemize}

 \name makes many correct predictions for sharp latency changes when flows compete for the available bandwidth after they starts. We acknowledge that some latency changes are fundamentally unpredictable (\eg start of a random flow) and \name can not predict those. Instead, \name focuses on learning interactions between competing flows on a shared link.

% \subsection{Component-wise analysis of \name's prediction quality}
\subsection{Component-wise analysis}
\label{eval:ss3}

\begin{figure}[t]
    \centering
    \includegraphics[width=.8\linewidth]{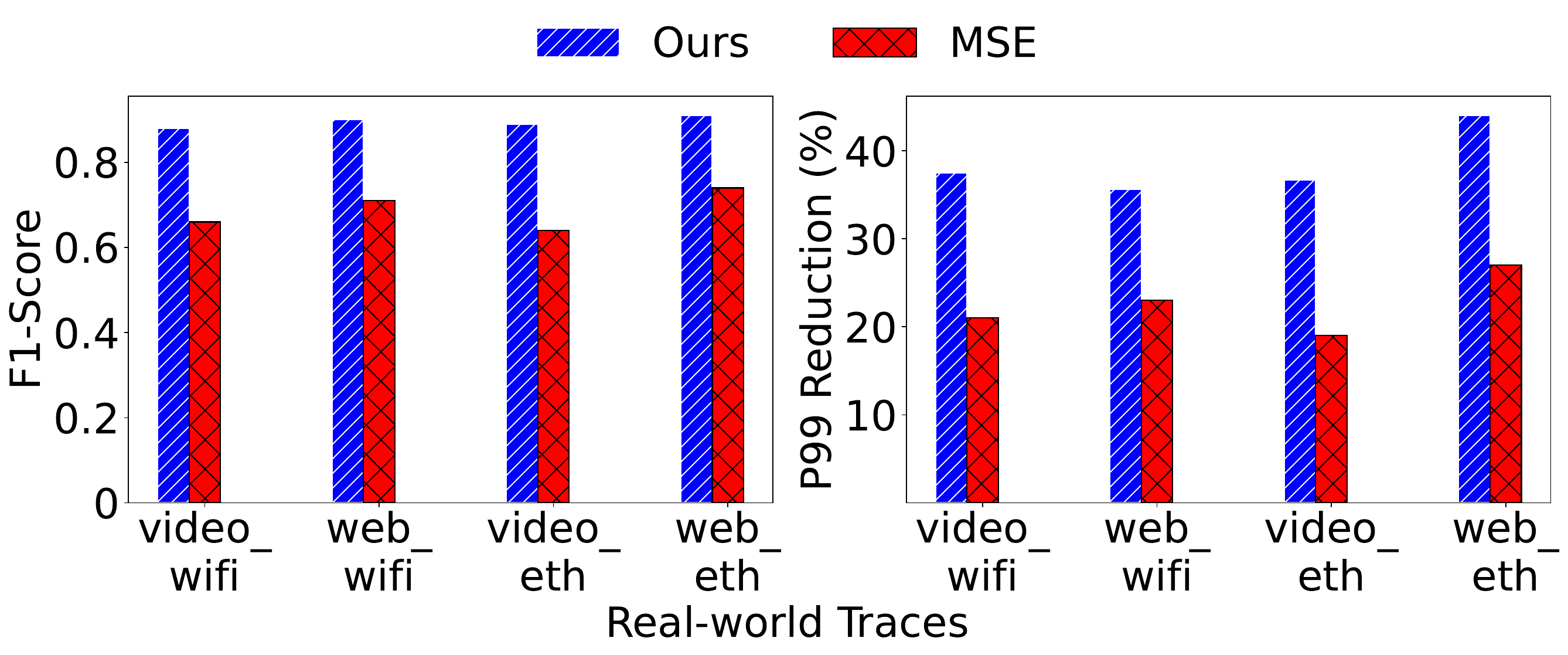}
    \caption{Compared with the standard loss function of MSE, training the Transformer with our custom loss function (\S\ref{sss:loss-function}) yields much higher prediction accuracy of sharp latency changes.
    % Prediction F1-score (precision + recall) for sharp changes is higher with our loss function 
    }
    \label{fig:mse_vs_lf}
\end{figure}

\paragraph*{Benefit of changing the loss function} We measure the effect of changing the loss function from the standard MSE to our change-point based loss function to focus on \emph{sharp changes} (introduced in \ref{ss:tf_control}). While the MSE is more optimal for learning values over the entire trend, it achieves a lower F1-score at actually predicting the sharp changes in latency. In Figure \ref{fig:mse_vs_lf} we see that \name achieves a higher F1-score at predicting these values across all of the real trace datasets by \emph{17-25\%}.

\begin{figure}[t]
    \centering
    \includegraphics[width=0.8\linewidth, keepaspectratio]
    {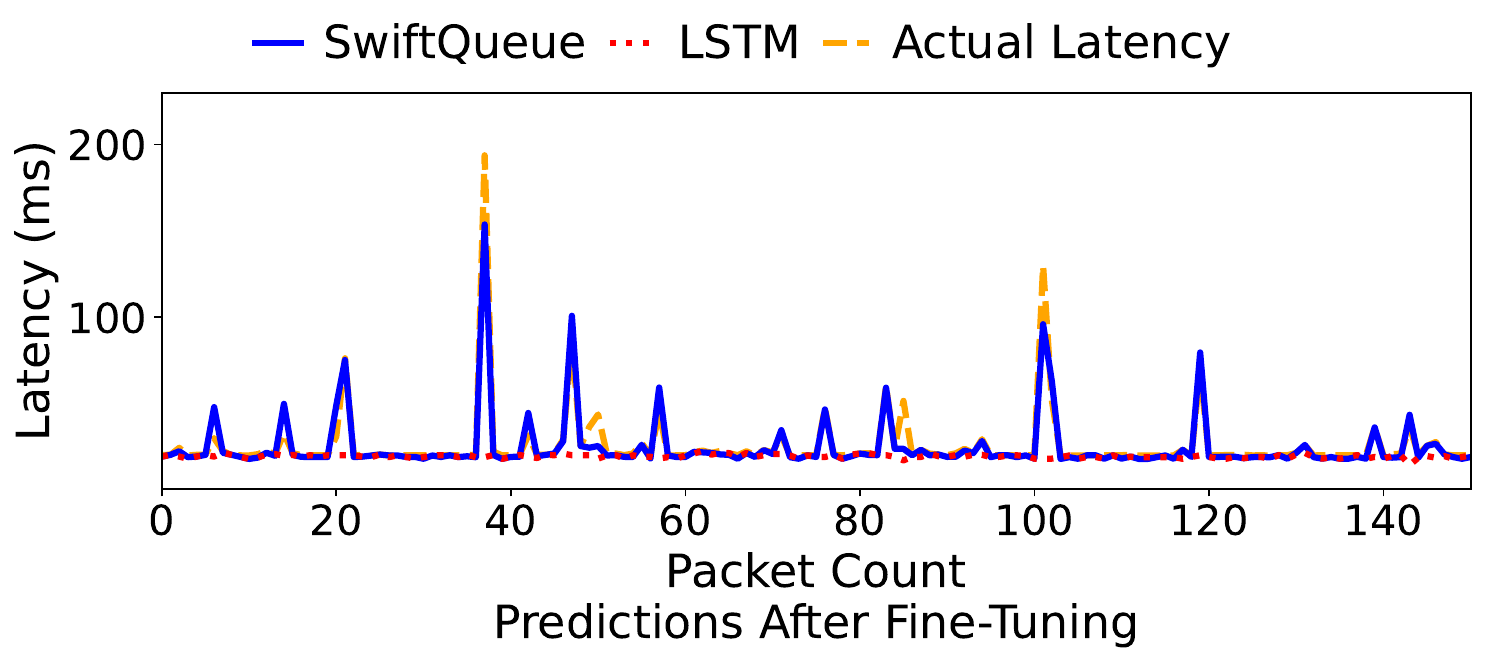}
    \caption{Visualizing the prediction of \name (with fine-tuning) on an example of 150 consecutive packets from the video\_wifi trace, compared with their actual latency values and the prediction of the most competitive baseline (LSTM).}
    % vs after only pre-training on a section of the video\_wifi trace}
    \label{fig:visualize_predictions}
\end{figure}

\begin{figure}[t]
    \centering
    \includegraphics[width=.75\linewidth]{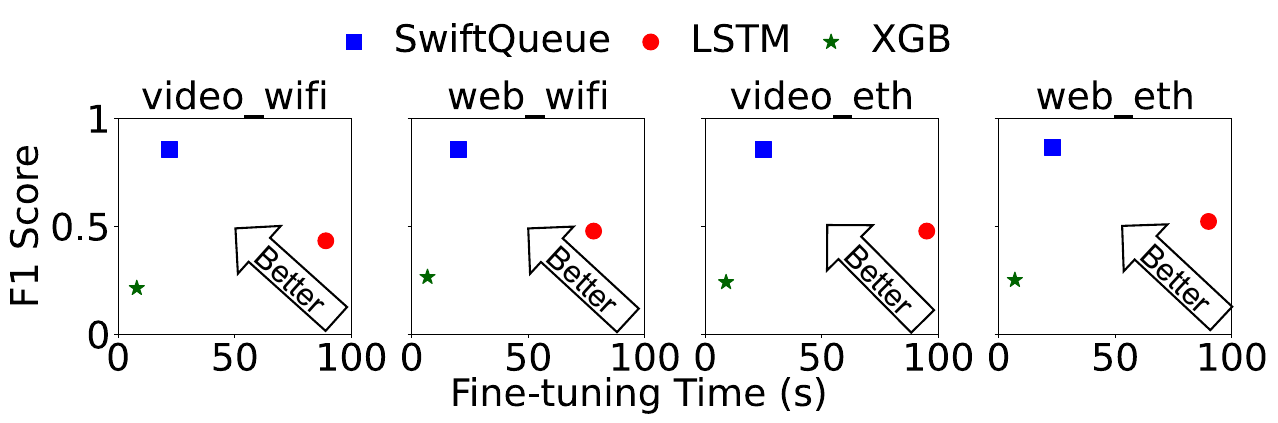}
    \caption{\name's prediction is more accurate (F1 score) than the baselines with a fine-tuning delay below 30 seconds. (LSTM and \name use the same hardware.) 
    % is faster to fine-tune compared to the LSTM baseline and has better prediction quality (higher F1-score on sharp changes)
    }
    \label{fig:f1_vs_ft_time}
\end{figure}

\paragraph*{Performance gains obtained from fine-tuning} Fine-tuning significantly helps improve the performance on \name. On real network traces, \name doesn't perform well without continuous fine-tuning. We observe this in Figure \ref{fig:visualize_predictions} where after fine-tuning, \name can capture both the \emph{sharp changes} in latency and also predict latency values which are closer to the \emph{true\_value}. We fine-tune the \emph{linear layers} of \name (and the LSTM). To measure the prediction quality, we measure the F1-score (combining precision and recall) on the predicted values of the sharp latency changes. \name is consistently better by \emph{45-65\%} times as compared to both baselines as seen in Figure \ref{fig:f1_vs_ft_time} in terms of prediction quality. \name is also significantly faster to fine-tune, as compared to the LSTM on the same single TITAN RTX 3080 GPU by \emph{3.8-4.4$\times$}. While fine-tuning the XGB model on the CPU is faster, the prediction quality is too low to be useful packet-level queue selection model.

\begin{figure}[t]
    \centering
    \includegraphics[width=.8\linewidth]{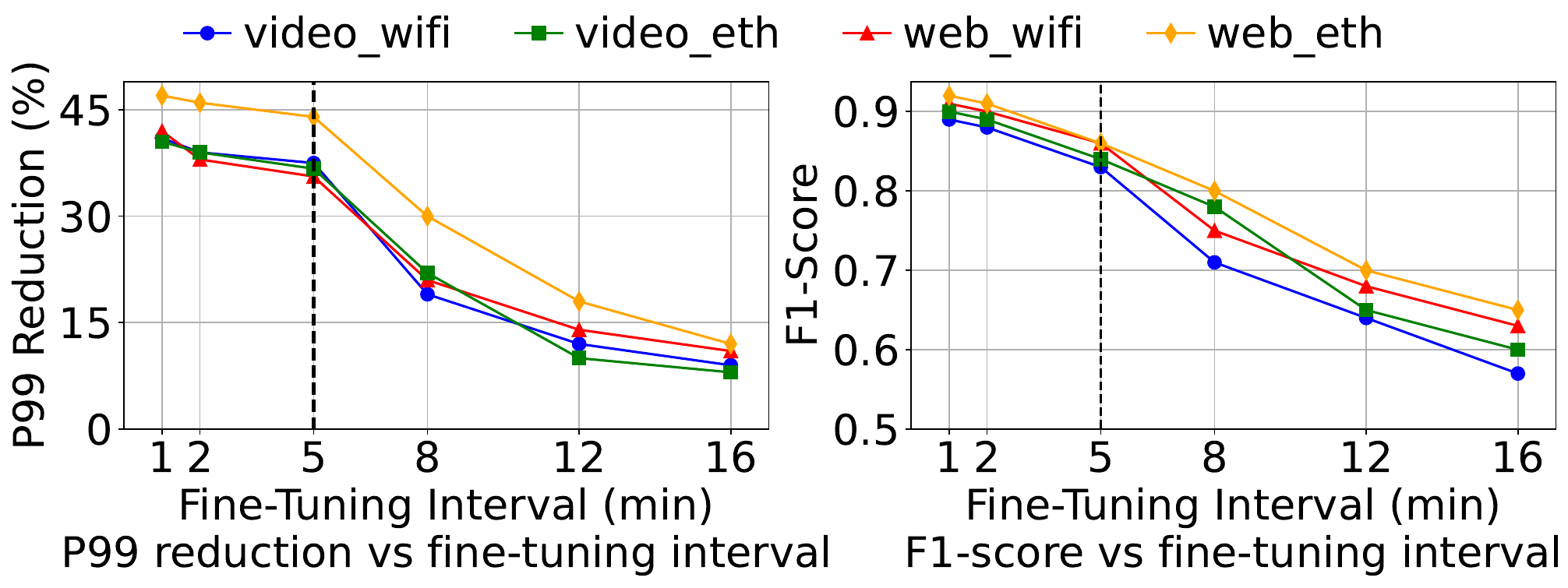}
    \caption{Latency reduction and prediction F1-score (precision + recall) generally drops with increasing fine-tuning interval, with the choice of fine-tuning interval of 5 minutes still achieving close-to-optimal latency reduction.}
    \label{fig:ft_frequency}
\end{figure}

\paragraph*{Effect of choosing the fine-tuning interval}
Choosing the right fine-tuning interval is important for \name to obtain the best prediction quality. While our fine-tuning speed for \name is fast, updating the model very frequently (e.g. every 10s of seconds) is computationally expensive (as we still use one GPU) and maybe unnecessary. In Figure \ref{fig:ft_frequency}, we measure the effect of the fine-tuning interval on the prediction accuracy. We see that fine-tuning at intervals of upto 5 minutes has negligible drop in prediction quality for our traces. Hence we don't need to use the GPU resource to fine-tune at a very short intervals if the gains are marginal.

\begin{figure}[t]
    \centering
    \includegraphics[width=.6\linewidth]{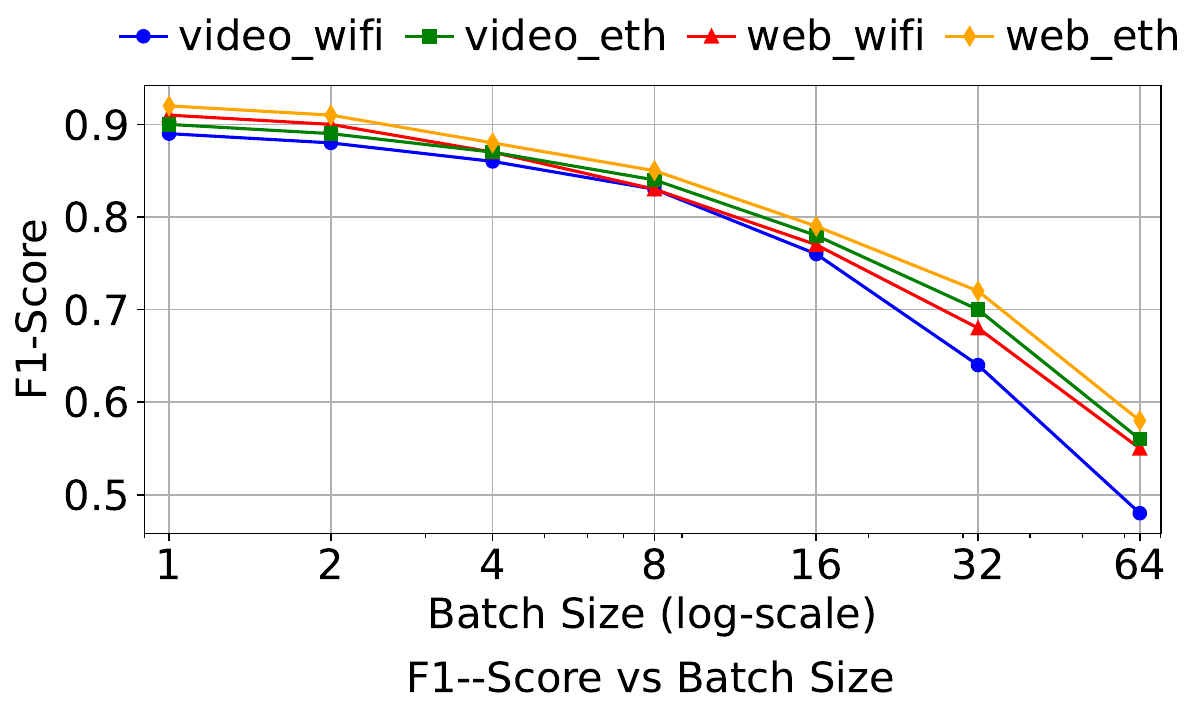}
    \caption{Prediction F1-score (precision + recall) drops with increasing batch size (more parallel prediction), with batch of 8 achieving a favorable tradeoff between parallel prediction and prediction accuracy.}
    \label{fig:ft_batch}
\end{figure}

\paragraph*{Batch size vs prediction quality}
We make \name's inference fast enough by predicting the next batch of latency values in order to make the decision in time before the next packet is sent out. We do not want to wait till we receive the previous latency as it might be too late. Increasing the prediction batch size makes it possible to predict the latency value a few packets earlier. However increasing the batch size comes at the cost of potentially reduced prediction quality. In Figure \ref{fig:ft_batch} we measure the prediction F1-score across increasing batch sizes and see that increasing the batch size excessively leads to a drop in prediction quality. Across our workloads a batch size of 8, allows for fast inference for ISP link speeds with a marginal drop in prediction quality.

\section{Discussion and Limitations}
\label{sec:future}

\textbf{\name for applications beyond L4S queue selection}
Currently L4S is chosen as an application as its \emph{static flow-level queue assignment} architecture can significantly benefit from a latency prediction model like \name, in order to meet its low-latency targets.
In addition, \name currently performs queue selection but does not modify the sender's TCP CC algorithm (L4S currently is limited to work with DCTCP/TCP Prague only).

We hope to inspire more research to optimize latency prediction and CC jointly. We see the potential to use the predictor to develop new proactive CC algorithms, which adapt to real-time network conditions beyond rule-based algorithms. We also envision future research with \name to extend to more applications like video-streaming, which can benefit from real-time network latency predictions for smoother use Quality-of-Experience.

\textbf{Excessively delayed ACKs, high traffic volume} If preceding ACKs are excessively delayed (\eg due to a high network propagation delay), \name cannot predict the next latency values accurately as it will not have enough past latency values to do so. If such a situation is detected (\eg using timeout windows), one solution for \name  might be to fall back to default L4S queue selection. Similarly, if the number of concurrent flows significantly scales up (\eg beyond $10000$s), the current \name Transformer architecture may also need to be scaled up to handle more complex inter-packet interactions.

\textbf{GPU hardware requirement and costs}
\name currently needs a GPU, albeit a cheap one (\S\ref{sec:implementation}), for low inference speed and may need a higher-end GPU to make in-time prediction for ultra-high bandwidth (\eg 10s of Gbps links). Running inference on GPUs does have additional costs and more research is needed to make \name go beyond L4S applications at consumer-degree ISPs, which has been the focus of this paper.

One method we envision for reducing GPU costs is integration with recent outcomes on network FPGA and hardware research which allows running ML models and making predictions at line-rate on low cost hardware~\cite{taurus, pegasus}. We leave the integration of \name with hardware beyond GPUs for future work.

Overall, while \name, as a first step, may not provide packet-level prediction and queue selection capabilities {\em universally across all networks}, it opens up new avenues of research for better Transformer-based packet-level analysis and optimizing more applications to benefit from it.
\section{Conclusion}
\label{sec:conclusion}

We present \name, which uses a Transformer for packet-level latency prediction and 
% . Using the predicted latency values, \name 
performs L4S-based packet-level queue selection to mitigate high tail latency. 
% , for packets with high predicted latency, for L4S flows. 
Using real traces and simulated ones, we show that \name outperforms all baselines at predicting sharp latency changes, and driven by the accurate latency prediction, \name achieves greater tail latency reduction for L4S flows. % with its prediction driven queue selection.

% \textbf{Ethical issues} This work does not raise any ethical issues.

%%
%% Bibliography
%%

%% Please use bibtex, 

\bibliography{main}

\appendix

\end{document}